\documentclass[11pt,a4paper]{article}
\pdfoutput=1

\usepackage{jcappub}
\usepackage{amsmath,amssymb,color,bm, graphicx}
\usepackage{enumerate,xstring, hyperref}

\def\be{\begin{equation}}
\def\ee{\end{equation}}
\def\bea{\begin{eqnarray}}
\def\eea{\end{eqnarray}}
\newcommand{\vs}{\nonumber\\}
\def\ba#1\ea{\begin{align}#1\end{align}}
\def\bg#1\eg{\begin{gather}#1\end{gather}}

\newcommand{\draw}{\leftarrow}

\newcommand{\fkonenl}{p\frac{P^\prime_m(p)}{P_m(p)}}
\newcommand{\fktwonl}{p^2\frac{P^{\prime\prime}_m(p)}{P_m(p)}}

\def\knl{k_\text{NL}}
\def\Ote{\hat{\varPi}}

\def\Plin{P_L}
\def\Pnl{P_m}

\renewcommand{\v}[1]{\bm{#1}}

\newcommand{\vx}{\v{x}}
\newcommand{\vv}{\v{v}}

\newcommand{\vr}{\v{r}}
\newcommand{\vk}{\v{k}}
\newcommand{\vq}{\v{q}}
\newcommand{\vp}{\v{p}}

\newcommand{\<}{\langle}
\renewcommand{\>}{\rangle}

\renewcommand{\d}{\delta}

\def\be{\begin{equation}}
\def\ee{\end{equation}}
\def\ben{\begin{eqnarray}}
\def\een{\end{eqnarray}}
\def\ba{\begin{array}}
\def\ea{\end{array}}

\def\ba#1\ea{\begin{align}#1\end{align}}

\newcommand{\bq}{\begin{eqnarray}}
\newcommand{\eq}{\end{eqnarray}}
\newcommand{\bes}{\begin{subequations}}
\newcommand{\ees}{\end{subequations}}

\def\R{\mathcal{R}}

\def\kunit{\:h\,{\rm Mpc}^{-1}}

\makeatletter
\newlength{\apb@width}
\newcommand{\autoparbox}[2][c]{\settowidth{\apb@width}{#2}\parbox[#1]{\apb@width}{#2}}
\newcommand{\includegraphicsbox}[2][]{\autoparbox{\includegraphics[#1]{#2}}}
\makeatother

\DeclareMathOperator{\cov}{Cov}

\newcommand{\comment}[1]{}

\begin{document}

\title{The squeezed matter bispectrum covariance with responses}

\author{Alexandre Barreira}
\emailAdd{barreira@MPA-Garching.MPG.DE}
\affiliation{Max-Planck-Institut f{\"u}r Astrophysik, Karl-Schwarzschild-Str.~1, 85741 Garching, Germany}

\abstract{We present a calculation of the angle-averaged squeezed matter bispectrum covariance $\cov\left(B_{m}(k_1, k_1', s_1), B_{m}(k_2, k_2', s_2)\right)$, $s_i \ll k_i,k_i'$ ($i=1,2$), that uses matter power spectrum responses to describe the coupling of large- to short-scale modes in the nonlinear regime. The covariance is given by a certain configuration of the 6-point function, which we show is dominated by response-type mode-coupling terms in the squeezed bispectrum limit. The terms that are not captured by responses are small, effectively rendering our calculation complete and predictive for linear $s_1,s_2$ values and any nonlinear values of $k_1,k_1',k_2,k_2'$. Our numerical results show that the squeezed bispectrum super-sample covariance is only a negligible contribution. We also compute the power spectrum-bispectrum cross-covariance using responses. Our derivation for the squeezed matter bispectrum is the starting point to calculate analytical covariances for more realistic galaxy clustering and weak-lensing applications. It can also be used in cross-checks of numerical ensemble estimates of the general bispectrum covariance, given that it is effectively noise-free and complete in the squeezed limit.}


\date{\today}

\maketitle
\flushbottom


\section{Introduction}\label{sec:intro}

The vast majority of parameter inference analyses using large-scale structure data are done using the 2-point matter correlation function, or in Fourier space, the power spectrum $P_m(k)$,
\bq\label{eq:Fourierdefs1}
(2\pi)^3 \Pnl(k_a) \delta_D(\vk_a + \vk_b) &=& \big<\tilde{\delta}(\vk_a)\tilde{\delta}(\vk_b)\big>,
\eq
where $k_a = |\vk_a|$, $\tilde{\delta}(\vk)$ is the Fourier transform of the three-dimensional matter density contrast and the angle brackets indicate an ensemble average. Measurements of the power spectrum are sufficient to describe the statistics of Gaussian random fields, but that is not the case of the late-time matter distribution in the Universe as nonlinear structure formation processes induce important non-Gaussian features. With cosmic time then, some of the information content available in the initial power spectrum has leaked to higher-order correlation functions, and including them in observational analysis allows us to recover some of that information. Next to the power spectrum, the simplest $N$-point function is the bispectrum ($N=3$), which is defined as
\bq\label{eq:Fourierdefs2}
(2\pi)^3 B_m(\vk_a, \vk_b, \vk_c) \delta_D(\vk_{abc}) &=& \big<\tilde{\delta}(\vk_a)\tilde{\delta}(\vk_b)\tilde{\delta}(\vk_c)\big>_c,
\eq
where the subscript $\big<\big>_c$ indicates it is a connected correlation function and $\vk_{abc} = \vk_a + \vk_b + \vk_c$ (we adopt this notation throughout). In addition to theoretical predictions for the power spectrum and bispectrum, parameter inference analyses using these statistics require also knowledge of the corresponding covariance and cross-covariance matrices. These include contributions up to the 6-point function, which are very involved mathematical quantities and makes estimating these covariance matrices very challenging, especially in the nonlinear regime of structure formation.

There are two main approaches to the evaluation of the covariance matrix of $N$-point functions. One is the {\it ensemble method}, in which one generates several statistically independent realizations of the density field using $N$-body simulations; the covariance matrix is then simply the covariance of the $N$-point functions measured in the ensemble. This approach yields a result that is valid on all scales probed by the size and resolution of the simulations, but at the cost of having to run many $N$-body simulations to yield sufficiently noise-free estimates (see e.g.~Refs.~\cite{2006PhRvD..74b3522S, 2012A&A...540A...9M, 2013MNRAS.429..344K, 2017MNRAS.465.1757G, 2017PhRvD..96b3528C, 2018arXiv180302132S, 2018arXiv180609499C} for examples of estimates of bispectrum covariance matrices with ensembles). The second approach is the {\it direct analytical calculation} of the higher-order $N$-point functions that specify the covariance matrix. This approach is practically noise-free and much less computationally intensive, but its accuracy is limited by the ability of current analytical methods to predict $N$-point functions in all regimes of structure formation. For instance, standard perturbation theory (SPT) methods \cite{Bernardeau/etal:2002} are relatively straightforward to implement, but the result is only valid on sufficiently large distance scales, $k \lesssim \knl \approx 0.3\ \kunit (z=0)$. The halo model of structure formation is another popular analytical approach, and although it is in principle predictive on all scales, it is also known to be somewhat inaccurate due to the simplifying assumptions behind it \cite{cooray/sheth, 2016PhRvD..93f3512S}. References \cite{2009A&A...508.1193J, 2013MNRAS.429..344K, 2018PhRvD..97d3532C, 2017PhRvD..96b3528C, 2018arXiv180206762D, 2018arXiv181207437R, 2018arXiv180602853G} are examples of works that undergo analytical evaluations of bispectra covariance matrices.

In this paper, we describe a novel analytical calculation of the bispectrum covariance based on power spectrum responses \cite{responses1}, which are functions that describe the {\it response} of the nonlinear matter power spectrum to the presence of long-wavelength density and tidal field perturbations. The responses can be measured efficiently in the nonlinear regime of structure formation using separate universe simulations \cite{li/hu/takada, 2014PhRvD..90j3530L, wagner/etal:2014, CFCpaper2, li/hu/takada:2016, lazeyras/etal, response, andreas}. In a perturbation theory sense, they describe the coupling of long- to short- wavelength modes, and quite importantly, they do so for nonlinear values of the short-wavelength modes; the response approach is thus an extension of SPT that is predictive in the nonlinear regime of structure formation. For example, with the response approach, the squeezed bispectrum  can be evaluated as
\bq\label{eq:intro1}
B_m(\vp, \vq, \vr) = \R_1(p, \mu_{\vp,\vr}) P_m(p)\Plin(r),\ \ \ \ \ r \ll p,q, \knl,
\eq
where $\R_1$ is called the first-order power spectrum response (measurable with separate universe simulations), $\mu_{\vp,\vr}$ is the cosine angle between $\vp, \vr$ and the subscripts $_m$ and $_L$ distinguish between the nonlinear and linear power spectrum, respectively. The only constraint on the validity of the above expression is that the long-wavelength mode $\vr$ must be in the perturbative regime; $\vp$ and $\vq$ can instead take on any nonlinear value. Conversely, the same calculation in SPT is only valid if all modes are in the perturbative regime $r,p,q \ll \knl$.

The usefulness of the response approach in the calculation of covariance matrices has already been demonstrated for the case of the power spectrum in Refs.~\cite{responses2, completessc, nocng}. In particular, in Ref.~\cite{nocng}, the authors have shown that the accuracy of lensing power spectra covariance matrices computed in the response approach may in fact be sufficient for parameter inference analysis using weak-lensing data from Euclid \cite{2011arXiv1110.3193L} and LSST \cite{2012arXiv1211.0310L}. This success can be traced back to the fact that the power spectrum covariance happens to be dominated by the squeezed mode-coupling interactions that responses describe in the nonlinear regime. In this paper, our goal is to demonstrate that the response approach is also a powerful tool in the evaluation of the covariance matrix of the bispectrum. As a first step towards that goal, we focus here on squeezed bispectrum configurations (cf.~Eq.~(\ref{eq:intro1})), for which (as we will see) the covariance can be readily evaluated with existing power spectrum response measurements from separate universe simulations. The response approach developed in Ref.~\cite{responses1} can be augmented to include also bispectrum response functions, which will allow to generalize the calculation presented here to more general bispectrum configurations.

\subsubsection*{Summary of the derivation}\label{eq:outline}

\begin{table}
  \centering
  \begin{tabular}{|c|c|c|c|c|c|c|c|c|c|c|}
    \hline
    \multicolumn{3}{|c|}{$P_m$ covariance} & \multicolumn{5}{|c|}{Squeezed $B_m$ covariance} & \multicolumn{3}{|c|}{Cross-covariance} \\
    \cline{1-11}
     \hline
     \hline  
    $PP$ & ${SSC}$ & $4pt_{nonSSC}$ & $PPP$ & $BB$ & $TP$ & ${SSC}$ & $6pt_{nonSSC}$& $BP$ & ${SSC}$ & $5pt_{nonSSC}$ \\
   \cline{1-11}
    $ \checkmark $ & $ \checkmark $ & $\checkmark^{*}$ & $ \checkmark $ & $ \checkmark $ & $ \checkmark $ & $ \checkmark $ & not incl. & $ \checkmark $ & $ \checkmark $ & not incl. \\
   \hline
  \end{tabular}
  \caption{Summary of the contributions to the covariance matrix of the matter power spectrum $P_m$, squeezed bispectrum $B_m$ and their cross-covariance, that we evaluate in this paper with the first- and second-order response functions (marked with $\checkmark$). The $4pt_{nonSSC}$ is marked with $\checkmark^{*}$ to highlight that responses capture the majority of this term, but that the calculation is strictly only complete in certain regimes. We do not explicitly calculate the $5pt,{nonSSC}$ and $6pt,{nonSSC}$ terms here, but their contribution can be added with standard perturbation theory, higher-order power spectrum responses and general bispectrum response functions (we argue however that the $6pt,{nonSSC}$ term is a small contribution).}
   \label{table:contsummary}
\end{table}

Despite being a straightforward derivation in the response approach, the calculation of the bispectrum covariance and power spectrum-bispectrum cross-covariance is forcibly an involved task to carry out, simply due to having to deal with terms up to the 6-point function. We thus provide here a summary of the calculation by listing all of the terms that we calculate with power spectrum responses. The busier reader can rely on this summarized account, skip Secs.~\ref{sec:responses}, \ref{sec:cov_PP}, \ref{sec:cov_BB} and \ref{sec:cov_BP}, and resume reading in Sec.~\ref{sec:results}, where we show a few numerical results.

The covariance of the power spectrum $\cov^{PP}$ is defined as
\bq\label{eq:covdefs1}
\cov^{PP}\big(k_1, k_2\big) &=& \big<\hat{P}_W(k_1)\hat{P}_W(k_2)\big> - \big<\hat{P}_W(k_1)\big>\big<\hat{P}_W(k_2)\big>,
\eq
the covariance of the bispectrum $\cov^{BB}$ is defined as
\bq\label{eq:covdefs2}
\cov^{BB}\big(k_1, k_1', s_1, k_2, k_2', s_2\big) &=& \big<\hat{B}_W(k_1, k_1', s_1)\hat{B}_W(k_2, k_2', s_2)\big> \nonumber \\ 
&&-  \big<\hat{B}_W(k_1, k_1', s_1)\big>\big<\hat{B}_W(k_2, k'_2, s_2)\big>
\eq
and the corresponding cross-covariance $\cov^{BP}$ is defined as
\bq\label{eq:covdefs2}
\cov^{BP}\big(k_1, k_1', s_1, k_2\big) &=& \big<\hat{B}_W(k_1, k_1', s_1)\hat{P}_W(k_2)\big> - \big<\hat{B}_W(k_1, k_1', s_1)\big>\big<\hat{P}_W(k_2)\big>,
\eq
where $\hat{P}_W(k_1)$ and $\hat{B}_W(k_1, k_1', s_1)$ are estimators of the power spectrum and bispectrum that we take to be
\bq\label{eq:estimatorsintro}
\hat{P}_W(k) &=& \frac{1}{V_WV_k} \int_k {\rm d}^3\vp\ \tilde{\delta}_W(\vp) \tilde{\delta}_W(-\vp), \\
\hat{B}_W(k_1, k_1', s_1) &=& \frac{1}{V_W V_{k_1k_1's_1}}  \int_{k_1}{\rm d}^3\vp\int_{k_1'}{\rm d}^3\vq \int_{s_1}{\rm d}^3\vr\ \tilde{\delta}_W(\vp)\tilde{\delta}_W(\vq)\tilde{\delta}_W(\vr) \delta_D(\vp+\vq+\vr), \nonumber \\
\eq
where $V_k = 4\pi k^2\Delta k$, $V_{k_1k_1's_1} = 8\pi^2 k_1k_1's_1\Delta k_1\Delta k_1'\Delta s_1$ and $\int_{k}{\rm d}^3\vp$ denotes averaging over a wavenumber shell with some width $\Delta k$ around $k$. The quantity $\tilde{\delta}_W$ denotes the density contrast modes observed inside some window function with volume $V_W$. Throughout, we always implicitly assume that the modes $k_1, k_1', s_1$ form a closed triangle, i.e., $k_1'^2 = {k_1^2 + s_1^2 + 2k_1s_1\mu_{k_1,s_1}}$, where $\mu_{k_1,s_1}$ is the cosine angle between the triangle sides $k_1$ and $s_1$. Further, $\vp, \vq$ will denote the integration modes associated with the hard (small-scale) modes of the angle-averaged triangle, i.e., $k_1, k_1'$, with $\vr$ being associated with the soft (large-scale) mode $s_1$. Correspondingly, we denote by $\vp', \vq', \vr'$ the integration modes of $k_2, k_2', s_2$, respectively.

Given that the estimator of the power spectrum involves the product of two Fourier modes, i.e. $\hat{P} \sim \tilde{\delta}\tilde{\delta}$, by Wick's theorem for a zero mean field, the covariance of the power spectrum will involve the product of two 2-point functions, as well as the connected 4-point function:
\bq\label{eq:covdefs3}
\cov^{PP} \supset \underbrace{\big<\tilde{\delta}\tilde{\delta}\big>\big<\tilde{\delta}\tilde{\delta}\big>}_{PP},  \underbrace{\big<\tilde{\delta}\tilde{\delta}\tilde{\delta}\tilde{\delta}\big>_c}_{4pt\ {\rm function}\ (SSC + nonSSC)},
\eq
where the underbraces indicate the names with which we refer to these terms in the derivation below; the 4-point function term is split into a super-sample covariance part (SSC) and the rest (non-SSC) of the contribution. Similarly, noting that $\hat{B} \sim \tilde{\delta}\tilde{\delta}\tilde{\delta}$ we have that
\bq
\label{eq:covdefs4}  \cov^{BB} &\supset& \underbrace{\big<\tilde{\delta}\tilde{\delta}\big>\big<\tilde{\delta}\tilde{\delta}\big>\big<\tilde{\delta}\tilde{\delta}\big>}_{PPP},\ \ \ \ \ \ \  \underbrace{\big<\tilde{\delta}\tilde{\delta}\tilde{\delta}\big>_c\big<\tilde{\delta}\tilde{\delta}\tilde{\delta}\big>_c}_{BB},\ \ \ \ \ \ \  \underbrace{\big<\tilde{\delta}\tilde{\delta}\tilde{\delta}\tilde{\delta}\big>_c\big<\tilde{\delta}\tilde{\delta}\big>}_{TP}, \underbrace{\big<\tilde{\delta}\tilde{\delta}\tilde{\delta}\tilde{\delta}\tilde{\delta}\tilde{\delta}\big>_c}_{6pt\ {\rm function}\ (SSC + nonSSC)} \\
\label{eq:covdefs5} \cov^{BP} &\supset& \underbrace{\big<\tilde{\delta}\tilde{\delta}\tilde{\delta}\big>_c\big<\tilde{\delta}\tilde{\delta}\big>}_{BP}, \underbrace{\big<\tilde{\delta}\tilde{\delta}\tilde{\delta}\tilde{\delta}\tilde{\delta}\big>_c}_{5pt\ {\rm function}\ (SSC + nonSSC)}.
\eq
Throughout, the superscripts in $\cov$ indicate which estimators we are taking the covariance of, and the subscripts indicate each of the above contributions; for example, $\cov^{BB}_{TP}$ denotes the $TP$ contribution to the bispectrum covariance. Note also that the decomposition of Eqs.~(\ref{eq:covdefs4}) and (\ref{eq:covdefs5}) holds for the general bispectrum and it is not peculiar to the squeezed limit.

\bigskip

In this paper, we show that, for the case of squeezed bispectrum configurations $s_1 \ll k_1, k_1', \knl$, the nonlinear matter power spectrum and corresponding first- and second-order response functions fully determine all of the above contributions, except the non-SSC part of the 5- and 6-point functions\footnote{Strictly, there is one permutation in $\cov^{BB}_{TP}$ that is also not completely given by responses, although the contribution that is left out is small (cf.~Sec.~\ref{sec:cov_BB_TP} below).}. We will argue that the $\cov^{BB}_{6pt,nonSSC}$ term is negligible, but that the $\cov^{BB}_{5pt,nonSSC}$ term could be important to keep the full squeezed bispectrum-power spectrum covariance matrix stable under inversion. The non-SSC part of the 4-point function has already been studied in Ref.~\cite{responses2}, in which the authors have shown that responses effectively account for the totality of the contribution if $k_1 \ll k_2$ (and vice-versa) in Eq.~(\ref{eq:covdefs1}), and about $70\%$ of the total for other mode configurations. Table \ref{table:contsummary} summarizes the terms derived in this paper.

\bigskip

An important aspect to stress is that the calculation derived in this paper based on the response approach is fully predictive in the nonlinear regime of the modes $k_i, k_i'$, with $s_i \ll  {\rm min}\{k_i, k_i', \knl\}$ (i = 1,2).

\bigskip

The outline of this paper is as follows. In Sec.~\ref{sec:responses}, we summarize the main aspects of the response approach to perturbation theory. Sections \ref{sec:cov_PP}, \ref{sec:cov_BB} and \ref{sec:cov_BP} display the derivation of the covariance of the matter power spectrum (which is a summary of past work \cite{responses1, responses2, completessc}),  matter bispectrum  and their cross-covariance, respectively. Section \ref{sec:results} displays a few numerical results of the covariance calculation, where we analyze in particular the relative size of the various contributions. Finally, we summarize and conclude in Sec.~\ref{sec:summary}. In appendix \ref{app:feynman}, we list the diagram rules for perturbation theory that we adopt in this paper. In appendix \ref{app:BBTPperms}, we display all of the mode permutations that constitute the $\cov^{BB}_{BP}$, $\cov^{BB}_{TP}$ and $\cov^{BP}_{BP}$ terms mentioned above. In appendix \ref{app:SSC_BB}, we derive with detail the SSC part of $\cov^{BB}$. In Appendix \ref{app:numericalevaluation} we describe the Monte Carlo integration scheme we use to obtain numerical results.

\section{Response approach to perturbation theory}\label{sec:responses}

In this section, we summarize the main concepts of the response approach to cosmological perturbation theory calculations. The interested reader can find in Ref.~\cite{responses1} a complete exposition of the formalism. Here, we limit ourselves to laying down the relevant equations and definitions that will be used in subsequent sections. Our diagram rules and conventions are listed in Appendix \ref{app:feynman}.

The $n$-th order matter power spectrum response $\R_n$ is defined with the following interaction vertex
\ba
&\lim_{\{r_a\} \to 0} \left(
\raisebox{-0.0cm}{\includegraphicsbox[scale=0.8]{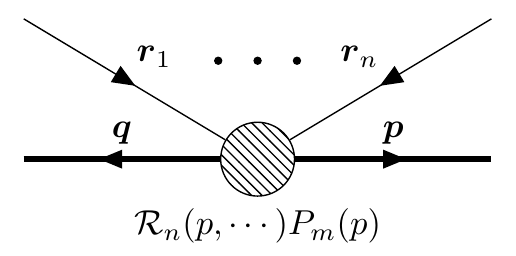}}
\right)   \nonumber \\
\nonumber \\
& = \frac12 \R_n(p;\, \{\mu_{\vp,\vr_a}\},\, \{\mu_{\vr_a,\vr_b}\},\, \{r_a/r_b\}) \Pnl(p) (2\pi)^3 \d_D(\vp+\vq- \vr_{12\cdots n})\,,
\label{eq:Rndef}
\ea
where the limit is interpreted as keeping the leading order term when the momenta $\vr_a$ are small in amplitude relative to $\vp$ and $\vq \approx -\vp$ (the subscripts $_a$, $_b$ label different soft modes). More precisely, we ignore corrections to the above equation that are of order $(r_a / p)^2$. The physical meaning of this response function is that it describes the {\it response} of the local nonlinear power spectrum $P_m(p)$ of the small-scale (hard) mode $\vp$ to the presence of $n$ long-wavelength (soft) modes $\vr_1, ..., \vr_n$. The dashed blob thus represents the fully evolved nonlinear matter power spectrum $\Pnl(p)$, as well as all its possible interactions (including loop interactions) with the $n$ long wavelength perturbations. The response $\R_n$ depends on the amplitude of the hard mode $p$, the cosine of the angle between the soft modes $\mu_{\vr_a,\vr_b}$, the cosine of the angle between the soft modes and the hard mode $\mu_{\vp,\vr_a}$, and the ratio of the amplitude of the soft modes $r_a/r_b$.

The diagrammatic representation of the response $\R_n$ facilitates understanding its link to the squeezed limit of the $(n+2)$-point matter correlation function. Concretely, {\it attaching} propagators (i.e., linear power spectra) to the soft momentum lines in Eq.~(\ref{eq:Rndef}) allows us to write
\ba
\lim_{\{r_a\}\to 0} \left(
\raisebox{-0.0cm}{\includegraphicsbox[scale=0.8]{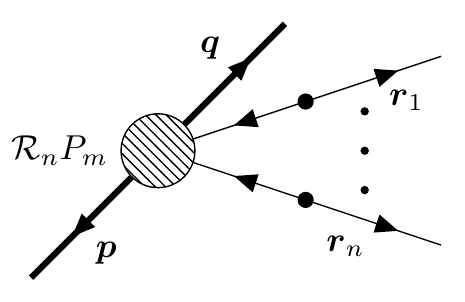}}
+ (\text{perm.}) \right)& = \<\tilde{\d}(\vp)\tilde{\d}(\vq)\tilde{\d}(\vr_1)\cdots \tilde{\d}(\vr_n)\>_{c, \R_n}
\vs
= n!\, \R_n(p;\, \{\mu_{\vp,\vr_a}\},\, \{\mu_{\vr_a,\vr_b}\},\, \{r_a/r_b\})
\Pnl(p) &\left[\prod_{a=1}^n\Plin(r_a)\right] \: (2\pi)^3 \d_D(\vp+\vq+\vr_{1\cdots n})\,,
\label{eq:sqnpt}
\ea
where the $n!$ factor accounts for the permutations of the $\vr_a$. The subscript ${}_{\R_n}$ in the $(n+2)$-connected correlator indicates that only certain contributions to the correlation function are actually captured by $\R_n$. The remaining contributions to $\<\tilde{\d}(\vp)\tilde{\d}(\vq)\tilde{\d}(\vr_1)\cdots \tilde{\d}(\vr_n)\>_c$ are either small in the squeezed limit, or are response-type terms as well, but described by lower order responses $\R_m$, $1\leq m < n$ and  perturbation theory kernels involving only the soft modes $\vr_a$. A concrete such example that will appear below in Sec.~\ref{sec:cov_BB_TP} is 
\bq
&\raisebox{-0.0cm}{\includegraphicsbox[scale=0.8]{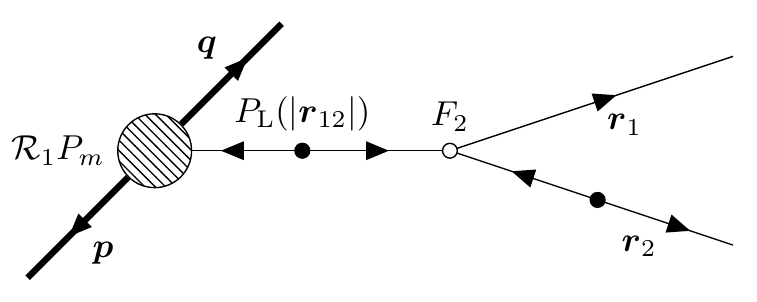}}
+\ (\vr_1 \leftrightarrow \vr_2) \vs
&\hspace*{2cm} = \R_1(p; \mu_{\vp,\vr_{12}}) P_m(p) \left[ 2 F_2(-\vr_{12},\vr_2) \Plin(|\vr_{12}|) \Plin(r_2)+ (\vr_1 \leftrightarrow \vr_2) \right] \vs
&\hspace*{2.5cm} \times (2\pi)^3 \d_D(\vp+\vq+\vr_{12}) \,.
\label{eq:sqTR1}
\eq

The local nonlinear matter power spectrum\footnote{The word "local" means that the power spectrum is meant to be measured in a finite volume whose typical size $L$ is sufficiently smaller than the wavelength of the soft modes, $L \ll 1/r_a$.} can be interpreted as a biased tracer of large-scale structure, and thus the $\R_n$ can be expanded in terms of all local gravitational observables (or operators $O$) associated with the $n$ long-wavelength modes \cite{biasreview}. These operators form a basis $\mathcal{K}_O$ that does not depend on the mode $k$ and that specify the angular dependence of $\R_n$:
\be
\R_n(p;\, \{\mu_{\vp,\vr_a}\},\, \{\mu_{\vr_a,\vr_b}\},\, \{r_a/r_b\})
= \sum_O R_O(p) \mathcal{K}_O^{(n)}(\{\mu_{\vp,\vr_a}\},\, \{\mu_{\vr_a,\vr_b}\},\, \{r_a/r_b\})\,.
\label{eq:Rndecomp}
\ee
The functions $R_O(p)$ are called \emph{response coefficients} and their physical meaning is that they describe the response of the power spectrum to the configuration of large-scale perturbations associated with the operator $O$. The scale-dependence of the coefficients can be worked out analytically at tree level in perturbation theory by plugging Eq.~(\ref{eq:Rndecomp}) into the tree-level expression of the $(n+2)$-point function in Eq.~(\ref{eq:sqnpt}). In the nonlinear regime of structure formation, the response coefficients can be evaluated using separate universe simulations \cite{li/hu/takada, 2014PhRvD..90j3530L, wagner/etal:2014, CFCpaper2, li/hu/takada:2016, lazeyras/etal, response, andreas} that simulate structure formation in the presence of long-wavelength perturbations.

For the calculation of the squeezed matter bispectrum covariance displayed in this paper we will need the two lowest order response functions. Specifically, $\R_1$ is given by
\bq
\label{eq:R1dec}  \R_1(p, \mu_1) = R_1(p) + R_K(p)\Big(\mu_1^2 -\frac13\Big),
\eq
and $\R_2$ by
\bq\label{eq:R2dec} 
&&\R_2(p; \mu_1,\mu_2,\mu_{12}, f_{12}) =  R_1(p)  \Bigg[{\frac{5}{7}} +  \frac{\mu_{12}}{2}\big(f_{12} + \frac{1}{f_{12}}\big)+ \frac27 \mu_{12}^2 \Bigg] \nonumber \\
&&+ R_K(p)\Bigg[\mu_1 \mu_2 \mu_{12} - \frac13 {\mu_{12}^2 } + \frac57 \left(\frac{( \mu_1 + f_{12} \mu_2)^2}{1 + f_{12}^2 + 2 f_{12} \mu_{12}} - \frac13 \right) (1 - \mu_{12}^2)  \nonumber \\ 
&& \qquad\qquad + \frac12  \mu_{12} \Bigg( \left(\mu_1^2 - \frac13\right)f_{12} + \left(\mu_2^2 - \frac13\right)\frac{1}{f_{12}} \Bigg) \Bigg] \nonumber \\
&& + \frac12 R_2(p) + \frac12 R_{K \d}(p) \Bigg[\mu_1^2 + \mu_2^2 - \frac23 \Bigg] + R_{K^2}(p) \Bigg[\mu_{12}^2 - \frac13\Bigg]  \nonumber \\
&& + R_{K.K}(p) \Bigg[\mu_1 \mu_2 \mu_{12} {-\frac13\mu_1^2-\frac13\mu_2^2+\frac19}\Bigg] + R_{KK}(p) \Bigg[\mu_1^2\mu_2^2 - \frac13\left(\mu_1^2 + \mu_2^2\right) + \frac19\Bigg] \nonumber \\ 
&& +  \frac32 R_{\Ote}(p) \left(\frac{( \mu_1 + f_{12} \mu_2)^2}{1 + f_{12}^2 + 2 f_{12} \mu_{12}} - \frac13 \right) (1 - \mu_{12}^2)\,,
\eq
where we have denoted $\mu_1 = \mu_{\vr_1, \vp}$, $\mu_2 = \mu_{\vr_2, \vp}$, $\mu_{12} = \mu_{\vr_1, \vr_2}$ and $f_{12} = r_1 / r_2$, for short.

In this paper, the 8 response coefficients that enter the above equations are evaluated as
\bq\label{eq:Ro} 
R_1(p) &=& 1 + G_1(p) - \frac13\fkonenl, \nonumber \\
R_K(p) &=& G_K(p) - \fkonenl, \nonumber \\
R_2(p) &=& \left(\frac{8}{21} G_1(p) + G_2(p)\right ) - \left(\frac{2}{9} + \frac23 G_1(p)\right)\fkonenl + \frac19\fktwonl - \frac23pG_1^{\prime}(p). \nonumber \\
R_{K\delta}(p) &=& \frac{1518}{1813}\left[\frac{8}{21}G_1(p) + G_2(p)\right] - \frac{41}{22}\left[\frac29 + \frac23G_1(p)\right]\fkonenl + \frac13\fktwonl \nonumber \\
R_{K^2}(p) &=& \frac{1}{21}G_1(p) - \frac{1}{6}\fkonenl,\nonumber \\ 
R_{K.K}(p) &=& -\frac{22}{13}G_1(p) +\frac32 \fkonenl, \nonumber \\
R_{KK}(p) &=& \frac{1476}{1813}\left[\frac{8}{21}G_1(p) + G_2(p)\right] - \frac{69}{44}\left[\frac29 + \frac23G_1(p)\right]\fkonenl + \frac12\fktwonl, \nonumber \\
R_{\Ote}(p) &=& -\frac{92}{273}G_1(p) +\frac13\fkonenl\,,
\eq
where a prime denotes a derivative w.r.t.~$p$. In the above equations, $G_1(p)$ and $G_2(p)$ are the isotropic growth-only response functions measured using separate universe simulations in Ref.~\cite{response}. The function $G_K(p)$ is the growth-only response to a tidal-field perturbation measured in Ref.~\cite{andreas} using a generalization of the separate universe technique to include long-wavelength tidal-field perturbations. 

\begin{figure}[t!]
        \centering
        \includegraphics[width=\textwidth]{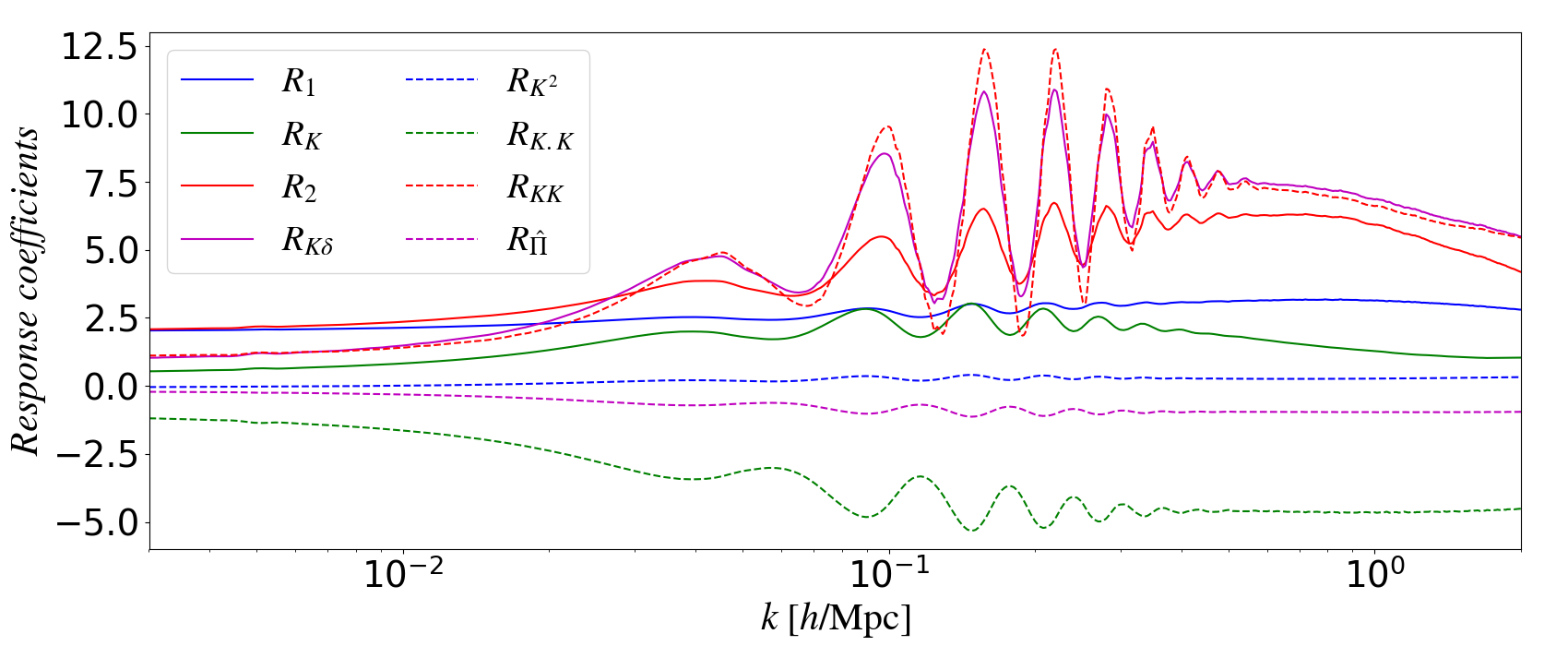}
        \caption{Scale-dependence of the 8 response coefficients that contribute to the full first- and second-order power spectrum response functions, $\R_1$ and $\R_2$, respectively (cf.~Eqs.~(\ref{eq:R1dec}), (\ref{eq:R2dec}) and (\ref{eq:Ro})). The curves shown correspond to redshift $z = 0$.}
\label{fig:Ro}
\end{figure}

The scale-dependence of the 8 response coefficients at redshift $z=0$ is shown in Fig.~\ref{fig:Ro}. It should be noted that only $R_1(p)$, $R_K(p)$ and $R_2(p)$ correspond strictly to the actual separate universe simulation measurements in the nonlinear regime. The remaining 5 coefficients have to date never been directly measured with simulations and the expressions shown above correspond to a physically motivated guess of their nonlinear shape that is based on the known relation between the nonlinear $R_1(p)$ and $R_2(p)$ coefficients and their tree-level limit (see Ref.~\cite{responses1} for the details). The good level of agreement between the ensemble method estimates of the power spectrum covariance of Ref.~\cite{blot2015} and the response-based calculation presented in Ref.~\cite{responses2} that uses the above expressions suggests, however, that the physically motivated guess is at least not drastically wrong.

\section{The matter power spectrum covariance}\label{sec:cov_PP}

This section presents the calculation of the matter power spectrum covariance in the response approach to perturbation theory. This was first derived in Refs.~\cite{responses1, responses2, completessc}, but here we repeat the main steps of the derivation for completeness and because it helps to build intuition for the more involved (although analogous) derivation of the squeezed bispectrum covariance in subsequent sections.

\subsection{Matter power spectrum estimator and covariance decomposition}\label{sec:cov_PP_estimator}

Let $\delta_W(\vx) = W(\vx)\delta(\vx)$ denote the three-dimensional matter density contrast field measured in some surveyed volume of the Universe described by a window function $W(\vx)$ that is unity inside the survey and zero outside. Its Fourier transform is given by (tildes indicate Fourier-space variables)
\bq\label{eq:dfourier}
\tilde{\delta}_W(\vp) =\int \frac{{\rm d}^3\vv}{(2\pi)^3}\tilde{W}(\vv) \tilde{\delta}(\vp-\vv).
\eq
We consider the following estimator of the angle-averaged power spectrum
\bq\label{eq:Pk3D}
\hat{P}_W(k) &=& \frac{1}{V_WV_k} \int_{k - \frac{\Delta_k}{2}}^{k + \frac{\Delta_k}{2}}\ p^2{\rm d}p \int_{0}^{2\pi}{\rm d}\varphi_{\vp}\int_{-1}^{1} {\rm d}{\mu}_{\vp} \  \tilde{\delta}_W(\vp) \tilde{\delta}_W(-\vp) \nonumber \\
&\equiv& \frac{1}{V_WV_k} \int_k {\rm d}^3\vp\ \tilde{\delta}_W(\vp) \tilde{\delta}_W(-\vp),
\eq
where $V_k = 4\pi k^2\Delta k$ is the Fourier integration volume ($\Delta k$ is the bin width), $V_W$ is the survey volume and the last equality defines our shorthand notation for $\int_k {\rm d}^3\vp$. This estimator is unbiased for scales sufficiently {\it inside} the survey, i.e.
\bq\label{eq:Pk3D<>}
\langle\hat{P}_W(k)\rangle &=& \frac{1}{V_WV_k} \int_k {\rm d}^3\vp \int\frac{{\rm d}^3\vv_1}{(2\pi)^3} \int\frac{{\rm d}^3\vv_2}{(2\pi)^3} \tilde{W}(\vv_1)\tilde{W}(\vv_2) \langle\tilde{\delta}(\vp-\vv_1)\tilde{\delta}(-\vp-\vv_2)\rangle\nonumber \\
&=& \frac{1}{V_WV_k} \int_k {\rm d}^3\vp \int\frac{{\rm d}^3\vv}{(2\pi)^3} |\tilde{W}(\vv)|^2 \Pnl(|\vp-\vv|) \nonumber \\
&\stackrel{p \gg v}{\approx}& \frac{1}{V_WV_k} \int_k {\rm d}^3\vp P_m(p) \int\frac{{\rm d}^3\vv}{(2\pi)^3} |\tilde{W}(\vv)|^2 \nonumber \\
&=& \frac{1}{V_k} \int_k {\rm d}^3\vp P_m(p) \nonumber \\
&=&  P_m(k),
\eq
where the approximation in the third equality follows from noting that the window function suppresses the integrand for $v \gg 1/V_W^{1/3}$, and hence, for modes $p \gg 1/V_W^{1/3}$ we can approximate $P_m(|\vp - \vv|) \approx P_m(p)$. In the steps above we have also used the following useful equation (which we will use throughout too)
\bq \label{eq:useful1} 
\tilde{W}(\vp) = \int \frac{{\rm d}^3\vv}{(2\pi)^3}\tilde{W}(\vv)\tilde{W}(\vp-\vv) = \left[\prod_{i=1}^n \int \frac{{\rm d}^3\vv_i}{(2\pi)^{3}}\tilde{W}(\vv_i)\right](2\pi)^3\delta_D(\vp - \vv_{12..n}),
\eq
which holds for the binary window functions we consider; note also that $\tilde{W}(0) \equiv V_W$.

The sample covariance of this power spectrum estimator can be written as 
\bq\label{covPP_def}
\cov^{PP}(k_1, k_2) &=& \langle\hat{P}_W(k_1)\hat{P}_W(k_2)\rangle - \langle\hat{P}_W(k_1)\rangle\langle\hat{P}_W(k_2)\rangle\nonumber \\
&=& \frac{1}{V_W^2V_{k_1}V_{k_2}} \int_{k_1} {\rm d}^3\vp \int_{k_2} {\rm d}^3\vp' \langle\tilde{\delta}_W(\vp) \tilde{\delta}_W(-\vp)\tilde{\delta}_W(\vp') \tilde{\delta}_W(-\vp')\rangle \nonumber \\
&-&\Pnl(k_1)\Pnl(k_2),
\eq
which shows that the key quantity to be evaluated is the 4-point function of the windowed density contrast $\tilde{\delta}_W$. Concretely, by Wick's theorem, the 4-point function of a zero-mean field gets contributions from the product of two 2-point functions and the connected 4-point function:
\bq\label{eq:wick4pt}
\langle\tilde{\delta}_W(\vp) \tilde{\delta}_W(-\vp)\tilde{\delta}_W(\vp') \tilde{\delta}_W(-\vp')\rangle &=& \langle\tilde{\delta}_W(\vp) \tilde{\delta}_W(-\vp)\rangle\langle\tilde{\delta}_W(\vp') \tilde{\delta}_W(-\vp')\rangle \nonumber \\
&+&\Big[ \langle\tilde{\delta}_W(\vp) \tilde{\delta}_W(\vp')\rangle\langle\tilde{\delta}_W(-\vp) \tilde{\delta}_W(-\vp')\rangle  + \left(\vp' \leftrightarrow -\vp' \right)\Big] \nonumber \\
&+& \langle\tilde{\delta}_W(\vp) \tilde{\delta}_W(-\vp)\tilde{\delta}_W(\vp') \tilde{\delta}_W(-\vp')\rangle_c 
\eq
Following the exact same steps and window function manipulations displayed in Appendix A of Ref.~\cite{completessc}, the above expression can be written as
\bq\label{eq:wick4pt_2}
\langle\tilde{\delta}_W(\vp) \tilde{\delta}_W(-\vp)\tilde{\delta}_W(\vp') \tilde{\delta}_W(-\vp')\rangle &=&  V_W^2 P_m(p)P_m(p') \nonumber \\
&+&\left[\Pnl(p)\right]^2 \Big[|\tilde{W}(\vp-\vp')|^2 + |\tilde{W}(\vp+\vp')|^2\Big] \nonumber \\
&+& \int \frac{{\rm d}^3\vv}{(2\pi)^3} |\tilde{W}(\vv)|^2 T_m(\vp, -\vp + \vv, \vp', -\vp' - \vv),
\eq
where the matter trispectrum $T_m$ is defined as 
\bq\label{eq:Tmdef}
(2\pi)^3 T_m(\vk_a, \vk_b, \vk_c, \vk_d) \delta_D({\vk_{abcd}}) = \langle\tilde{\delta}(\vk_a)\tilde{\delta}(\vk_b)\tilde{\delta}(\vk_c)\tilde{\delta}(\vk_d)\rangle_c.
\eq
The contribution from the first term on the right-hand side of Eq.~(\ref{eq:wick4pt_2}) trivially cancels out with the term $P_m(k_1)P_m(k_2)$ in Eq.~(\ref{covPP_def}). The second and third terms yield the so-called {\it Gaussian} and {\it non-Gaussian} terms, respectively. This is a terminology that has been used in a few matter-power-spectrum-covariance related papers. Here, to keep the notation consistent with the bispectrum covariance contributions, we refer to these terms as the $PP$ and 4-point function terms, respectively. We describe each of these in turn next.

\subsection{The $PP$ term}\label{sec:cov_PP_G}

Plugging the second term on the right-hand side of Eq.~(\ref{eq:wick4pt_2}) into Eq.~(\ref{covPP_def}) yields
\bq\label{eq:Gterm}
\cov^{PP}_{PP}(k_1, k_2) &=& \frac{1}{V_W^2V_{k_1}V_{k_2}} \int_{k_1} {\rm d}^3\vp \int_{k_2} {\rm d}^3\vp' \left[\Pnl(p)\right]^2 \Big[|\tilde{W}(\vp-\vp')|^2 + |\tilde{W}(\vp+\vp')|^2\Big] \nonumber \\
&\approx& \frac{(2\pi)^6}{V_W^2V_{k_1}V_{k_2}} \int_{k_1} {\rm d}^3\vp \int_{k_2} {\rm d}^3\vp' \left[\Pnl(p)\right]^2 \Big[\delta_D(\vp-\vp')\delta_D(-\vp+\vp') \nonumber \\
&& \ \ \ \ \ \ \ \ \ \ \ \ \ \ \ \ \ \ \ \ \ \ \ \ \ \ \ \ \ \ \ \ \ \ \ \ \ \ \ \ \ \ \ \ \ + \delta_D(\vp+\vp')\delta_D(-\vp-\vp')\Big], \nonumber \\
\eq
where in the second equality we have approximated $\tilde{W}(\vp + \vp') \approx (2\pi)^3\delta_D(\vp + \vp')$, which is a decent approximation\footnote{More precisely, $\tilde{W}(\vp-\vp')$ constraints $\vp$ and $\vp'$ to be the same up to a correction of size $1/V_W^{1/3}$, which is small if we restrict ourselves to modes sufficiently deep inside the survey. Hence, $|\tilde{W}(\vp - \vp')|^2 = \tilde{W}(\vp - \vp')\tilde{W}(-\vp + \vp') \approx (2\pi)^6\delta_D(\vp-\vp')\delta_D(-\vp+\vp')$.} if we are interested in modes $p, p' \gg 1/V_W^{1/3}$. Carrying out one of the integrals yields a factor of 2 from the two sets of Dirac delta functions with a constraint that $k_1$ and $k_2$ must be in the same wavenumber bin, as well as a factor of $\delta_D(0) \equiv V_W/(2\pi)^3$,
\bq\label{eq:Gterm_2}
\cov^{PP}_{PP}(k_1, k_2) &=& \delta_{k_1k_2}\frac{2\ (2\pi)^3}{V_WV_{k_1}V_{k_2}} \int_{k_1} {\rm d}^3\vp \left[\Pnl(p)\right]^2 \nonumber \\
&\approx& \delta_{k_1k_2}\frac{2\ (2\pi)^3}{V_WV_{k_1}}\left[\Pnl(k_1)\right]^2,
\eq
where the approximation follows from assuming sufficiently narrow bin widths that allow the power spectrum to be taken out of the integral.

\subsection{The connected 4-point function term}\label{sec:cov_PP_NG}

The 4-point function contribution is determined by a certain configuration of the matter trispectrum as \cite{1999ApJ...527....1S, 2006MNRAS.371.1188H, takada/hu:2013}
\bq\label{eq:NGterm}
\cov^{PP}_{4pt}(k_1, k_2) &=& \frac{1}{V_W^2V_{k_1}V_{k_2}} \int_{k_1} {\rm d}^3\vp \int_{k_2} {\rm d}^3\vp' \int \frac{{\rm d}^3\vv}{(2\pi)^3} |\tilde{W}(\vv)|^2 T_m(\vp, -\vp + \vv, \vp', -\vp' - \vv). \nonumber \\
\eq
In perturbation theory \cite{Bernardeau/etal:2002}, the trispectrum can be expanded into terms that contribute at different loop orders
\bq\label{eq:Tmexp}
T_m(\vp, -\vp + \vv, \vp', -\vp' - \vv) = T_m^{\rm tree} + T_m^{\rm 1-loop} + T_m^{\rm 2-loop} + \cdots,
\eq
and one can further identify two physically distinct contributions to the above terms. One is called {\it super-sample covariance} (SSC) \cite{takada/hu:2013, completessc} and it comprises all the terms/diagrams that enter Eq.~(\ref{eq:Tmexp}) that are zero if $\vv = 0$. More precisely, we can write
\bq\label{eq:Tssc}
T^{\rm SSC}(\vp, -\vp, \vp', -\vp'; \vv) = \left[ \lim_{v \to 0} \frac{\partial}{\partial [\Plin(v)]} T_m(\vp, -\vp + \vv, \vp', -\vp' - \vv) \right] \Plin(v).
\eq
Physically, this term describes the coupling between measured nonlinear sub-survey modes with unobserved super-survey Fourier modes, i.e.,  modes with wavelengths larger than $V_W^{1/3}$. 

The other term has been called {\it connected non-Gaussian} term in past literature and it corresponds to the rest of the contribution, i.e., all of the terms in Eq.~(\ref{eq:Tmexp}) that are non-zero when $\vv = 0$: $T_m^{nonSSC}(\vp, -\vp, \vp', -\vp') = T_m(\vp, -\vp + \vv, \vp', -\vp' - \vv) - T^{\rm SSC}(\vp, -\vp, \vp', -\vp'; \vv)$. This term describes the coupling of sub- to sub-survey modes that is induced by nonlinear structure formation (it is present at all times, however, in cosmologies with primoridal non-Gaussianity). Strictly speaking, $T_m^{nonSSC}$ also depends on the window function momenta $\vv$, but that dependence can be ignored if $p, p' \gg 1/V_W^{1/3}$. Equation (\ref{eq:NGterm}) can thus be split into two as
\bq\label{eq:NGterm2}
\cov^{PP}_{4pt}(k_1, k_2) = \cov^{PP}_{4pt, nonSSC}(k_1, k_2) + \cov^{PP}_{4pt, SSC}(k_1, k_2),
\eq
with
\bq
\label{eq:cNGterm} \cov^{PP}_{4pt, nonSSC}(k_1, k_2) &=& \frac{1}{V_W V_{k_1}V_{k_2}} \int_{k_1} {\rm d}^3\vp \int_{k_2} {\rm d}^3\vp' T_m^{nonSSC}(\vp, -\vp, \vp', -\vp'), \\
\label{eq:SSCterm} \cov^{PP}_{4pt, SSC}(k_1, k_2) &=& \frac{1}{V_W^2 V_{k_1}V_{k_2}} \int_{k_1} {\rm d}^3\vp \int_{k_2} {\rm d}^3\vp' \int \frac{{\rm d}^3\vv}{(2\pi)^3} |\tilde{W}(\vv)|^2 T^{\rm SSC}(\vp, -\vp, \vp', -\vp'; \vv). \nonumber \\
\eq
The trispectra terms that enter the above two equations are what can be evaluated with the response approach to perturbation theory.

\subsubsection{The non-SSC contribution}\label{sec:cov_PP_cNG}

Here, we display the calculation of $T_m^{nonSSC}(\vp, -\vp, \vp', -\vp')$ presented in Ref.~\cite{responses2}, which includes the totality of the tree-level contribution and part of the 1-loop term.

The tree-level result can be written as the {\it stitching} of two results
\bq\label{eq:stitched}
T_m^{nonSSC, st-tree}(\vp, -\vp, \vp', -\vp') =
\begin{cases} 
T_m^{nonSSC, SPT-tree}(\vp, -\vp, \vp', -\vp') &,\  {\rm if}\ \frac{p_{\rm soft}}{p_{\rm hard}} > f_\text{sq}\\ 
T_m^{nonSSC, \R_2-tree}(\vp, -\vp, \vp', -\vp')&, \ {\rm otherwise}
\end{cases}\;, \nonumber \\
\eq
where
\bq\label{eq:TmcngR2}
T_m^{nonSSC, \R_2-tree}(\vp, -\vp, \vp', -\vp') = 2 \R_2(p_{\rm hard}, \mu_{\vp,\vp'}, -\mu_{\vp,\vp'}, -1, 1) \Pnl(p_{\rm hard}) \left[\Plin(p_{\rm soft})\right]^2, \nonumber \\
\eq
$T_m^{nonSSC, SPT-tree}$ is the tree-level trispectrum as given by standard perturbation theory \cite{1999ApJ...527....1S}, $p_{\rm hard} = {\rm max}\{p, p'\}$ and $p_{\rm soft} = {\rm min}\{p, p'\}$. The parameter $f_\text{sq}$ defines when the configuration is considered squeezed and when to use which branch. Specifically, when $\vp$ and $\vp'$ have approximately the same amplitude, then one uses SPT to evaluate the trispectrum (upper branch of Eq.~(\ref{eq:stitched})); this branch is only predictive if $p_{\rm soft}, p_{\rm hard} \ll \knl$, where $\knl$ is the nonlinear scale. On the other hand, if one of the modes is sufficiently harder than the other, then we are in squeezed configurations, and hence, one can use the response approach (lower branch of Eq.~(\ref{eq:stitched}), obtained with the $n=2$ case of Eq.~(\ref{eq:sqnpt})). In this lower branch, $p_{\rm hard} > p_{\rm soft}/f_{sq}$, the result is predictive for any nonlinear value of the hard mode. In our numerical results, we consider $f_\text{sq} = 0.5$; appendix D of Ref.~\cite{responses2} illustrates that this choice ensures a sufficiently smooth transition between the two branches in the regime where they overlap: $p_{\rm hard}, p_{\rm soft} \ll \knl$. The interested reader can find more details on the evaluation of Eq.~(\ref{eq:stitched}) in Sec.~3 of Ref.~\cite{responses2}.

The 1-loop contribution to $T_m^{nonSSC}(\vp, -\vp, \vp', -\vp')$ that is captured by response functions is given by the following diagram
\ba\label{eq:Tm_cNG_loop}
& T_m^{nonSSC, \R_2-1loop}(\vp, -\vp, \vp', -\vp') = \nonumber \\
&   = \raisebox{-0.0cm}{\includegraphicsbox[scale=0.8]{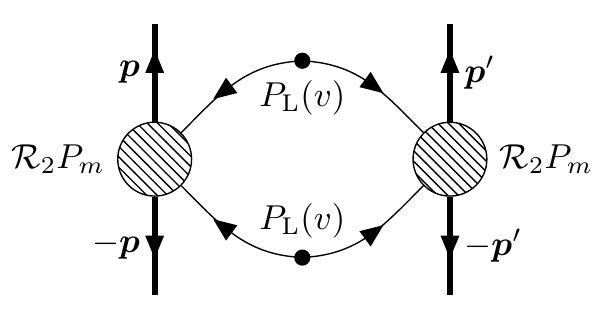}}
+(\text{perm.}) = \nonumber \\ 
& = 2 \int \frac{{\rm d}^3\vv}{(2\pi)^3}[\Plin(v)]^2 \R_2(p,\mu_{\vp\vv},-\mu_{\vp\vv},-1,1)\ \R_2(p',\mu_{\vp'\vv},-\mu_{\vp'\vv},-1,1) \Pnl(p)\Pnl(p'). \nonumber \\ 
\ea
In the integral over the amplitude of $\vv$ we impose a cutoff at $v_{\rm max} = {\rm min}\{p_{\rm soft}, \knl\}$. In addition to the contribution written above, the 1-loop term contains also contributions from a number of diagrams that cannot be captured by responses, as well as the diagrams captured by Eq.~(\ref{eq:Tm_cNG_loop}) for $v > v_{\rm max}$. These can be added to the calculation by following a similar (but more involved) stitching procedure as that used at tree-level (cf.~Eq.~(\ref{eq:stitched})).

Overall, we compute $\cov^{PP}_{4pt, nonSSC}(k_1, k_2)$ using Eq.~(\ref{eq:cNGterm}) with 
\bq\label{eq:Tm_cNG_2}
T_m^{nonSSC}(\vp, -\vp, \vp', -\vp') = T_m^{nonSSC, st-tree}(\vp, -\vp, \vp', -\vp') + T_m^{nonSSC, \R_2-1loop}(\vp, -\vp, \vp', -\vp'). \nonumber \\
\eq
In Ref.~\cite{responses2}, this calculation was compared to the numerical estimates of Ref.~\cite{blot2015} based on an ensemble of $> 12000$ simulations. These response- and ensemble-based calculations were shown to be in agreement in squeezed configurations, i.e., ${\rm min}\{k_1, k_2\} \ll {\rm max}\{k_1, k_2\}$, which is the regime in which the response calculation is expected to be virtually complete. In regimes when $k_1, k_2$ have comparable amplitudes the response-based result, underestimates the simulation results by $\approx 30\%$; the inclusion of higher-loop terms in Eq.~(\ref{eq:Tmexp}), as well as the inclusion of the rest of the 1-loop term that cannot be described with responses can however improve the accuracy of the calculation.

\subsubsection{The SSC contribution}\label{sec:cov_PP_SSC}

The calculation of $T^{\rm SSC}(\vp, -\vp, \vp', -\vp'; \vv)$ with the response approach was presented in Ref.~\cite{completessc}, and it is given by (see also Refs.~\cite{takada/hu:2013, li/hu/takada, akitsu/takada/li})
\bq\label{eq:sscdiagrams}
T^{\rm SSC}(\vp, -\vp, \vp', -\vp'; \vv) &=& \raisebox{-0.0cm}{\includegraphicsbox[scale=0.85]{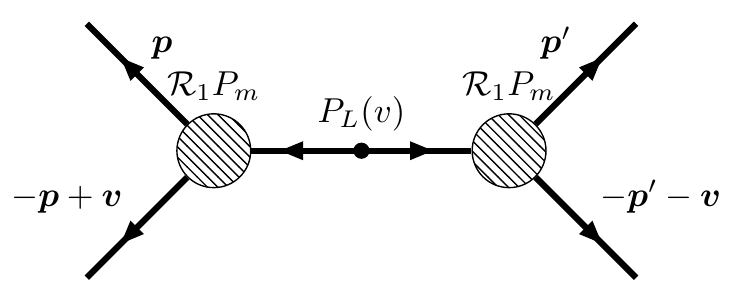}} \nonumber \\
&=& \R_1(p, -\mu_{\vp,\vv}) \R_1(p', \mu_{\vp',\vv})\Pnl(p)\Pnl(p')\Plin(v).
\eq
This expression is valid for nonlinear amplitudes of $\vp$, $\vp'$ and linear $\vv$. The sizes of current and future surveys are however sufficiently large to ensure that the contribution from nonlinear $\vv$ gets suppressed by $|\tilde{W}(\vv)|^2$ in Eq.~(\ref{eq:SSCterm}). Equations (\ref{eq:SSCterm}) and (\ref{eq:sscdiagrams}) thus effectively capture the totality of the SSC contribution to the matter power spectrum covariance.

\section{The matter bispectrum covariance}\label{sec:cov_BB}

Having warmed up with the case of the power spectrum covariance in the last section, we now turn to the calculation of the squeezed matter bispectrum covariance. The main idea and derivation steps are analogous to those of the power spectrum, but the calculation is naturally more involved by virtue of dealing with higher-order correlation functions.

\subsection{Matter squeezed bispectrum estimator and covariance decomposition}\label{sec:cov_PP_estimator}

We work with the following estimator of the matter bispectrum 
\bq\label{eq:sqB_estimator}
\hat{B}_W(k_1, k_1', s_1) = \frac{1}{V_WV_{k_1k_1's_1}} \int_{k_1}{\rm d}^3\vp\int_{k_1'}{\rm d}^3\vq \int_{s_1}{\rm d}^3\vr\ \tilde{\delta}_W(\vp)\tilde{\delta}_W(\vq)\tilde{\delta}_W(\vr)\ \delta_D(\vp+\vq+\vr), \nonumber \\
\eq
where $V_{k_1k_1's_1} = 8\pi^2 k_1k_1's_1\Delta k_1\Delta k_1'\Delta s_1$. In our notation, we always implicitly assume that the sizes of $\{k_1, k_1', s_1\}$ form a closed squeezed triangle with $k_1, k_1' \gg s_1$. Analogously to the case of the power spectrum estimator in Eq.~(\ref{eq:Pk3D}), this estimator can be shown to be unbiased for modes sufficiently inside the survey ($k_1, k_1',s_1 \gg 1/V_W^{1/3}$), i.e., 
\bq\label{eq:sqB_estimator_2}
\Big< \hat{B}_W(k_1, k_1', s_1)\Big> &=& B_m(k_1, k'_1, s_1) \nonumber \\
&=& \frac{1}{V_{k_1k_1's_1}}  \int_{k_1}{\rm d}^3\vp\int_{k_1'}{\rm d}^3\vq \int_{s_1}{\rm d}^3\vr\ B_m(\vp, \vq, \vr) \delta_D(\vp+\vq+\vr) \nonumber \\
&\stackrel{k_1,k_1' \gg s_1}{=}& \frac{1}{V_{k_1k_1's_1}}  \int_{k_1}{\rm d}^3\vp\int_{k_1'}{\rm d}^3\vq \int_{s_1}{\rm d}^3\vr\ \left[R_1(p) + R_K(p)\left(\mu_{\vp,\vr}^2 - \frac{1}{3}\right)\right] \nonumber \\
&& \Pnl(p)\Plin(r)\delta_D(\vp+\vq+\vr),
\eq
where the second equality writes the squeezed bispectrum in terms of the first order power spectrum response (cf.~Eq.~(\ref{eq:intro1}) and Sec.~\ref{sec:responses})\footnote{We note in passing that the bin-averaged squeezed bispectrum of Eq.~(\ref{eq:sqB_estimator_2}) depends explicitly on the tidal response $R_K$, but the angle-averaged squeezed bispectrum definition of Ref.~\cite{response} does not.  This has to do with the fact that in Ref.~\cite{response}, the amplitude of the momenta $\vp$, $\vq$ is not constrained to be inside a given wavenumber bin, and hence, during the angle average, $\mu_{\vp,\vr}$ varies freely from $-1$ to $1$ and the $R_K$ contribution cancels out. On the other hand, for the case of averages over sufficiently narrow bins, then $\mu_{\vp,\vr}$ is constrained to be approximately equal to the cosine angle between the $k_1$ and $s_1$ sides of the triangle, which leads to the $R_K$ contributing in general to the bin-averaged bispectrum.}.

The covariance of the estimator of Eq.~(\ref{eq:sqB_estimator}) can be written as
\bq\label{eq:covBB_def}
\cov^{BB} &\equiv& \cov^{BB}(k_1, k_1', s_1, k_2, k_2', s_2) \nonumber \\
&=& \Big<\hat{B}_W(k_1, k_1', s_1)\hat{B}_W(k_2, k_2', s_2)\Big> - \Big<\hat{B}_W(k_1, k_1', s_1)\Big>\Big<\hat{B}_W(k_2, k_2', s_2)\Big> \nonumber \\
&=& \frac{1}{V_W^2V_{k_1k_1's_1}V_{k_2k_2's_2}} \int_{k_1}{\rm d}^3\vp\int_{k_1'}{\rm d}^3\vq \int_{s_1}{\rm d}^3\vr \int_{k_2}{\rm d}^3\vp'\int_{k_2'}{\rm d}^3\vq' \int_{s_2}{\rm d}^3\vr' \nonumber \\
&& \delta_D(\vp+\vq+\vr)\delta_D(\vp'+\vq'+\vr') \Big< \tilde{\delta}_W(\vp) \tilde{\delta}_W(\vq) \tilde{\delta}_W(\vr) \tilde{\delta}_W(\vp') \tilde{\delta}_W(\vq') \tilde{\delta}_W(\vr') \Big> \nonumber \\
&-& B_m(k_1, k_1', s_1)B_m(k_2, k_2', s_2).
\eq
The calculation of the bispectrum covariance thus boils down to the evaluation of the 6-point correlation function of the windowed density field, which can be split into four distinct types of contributions \cite{2006PhRvD..74b3522S, 2013MNRAS.429..344K}:
\begin{enumerate}
\item one given by the product of three 2-point correlation functions $\big<\tilde{\delta}\tilde{\delta}\big>\big<\tilde{\delta}\tilde{\delta}\big>\big<\tilde{\delta}\tilde{\delta}\big>$, which we refer to as the $PPP$ term; 

\item another given by the product of two connected 3-point correlation functions $\big<\tilde{\delta}\tilde{\delta}\tilde{\delta}\big>_c\big<\tilde{\delta}\tilde{\delta}\tilde{\delta}\big>_c$, which we call the $BB$ term; 

\item one given by the product of a connected 4-point function with a 2-point function $\big<\tilde{\delta}\tilde{\delta}\tilde{\delta}\tilde{\delta}\big>_c\big<\tilde{\delta}\tilde{\delta}\big>$, which we call the $TP$ term; 

\item and finally, one given by the contribution of the connected 6-point correlation function $\big<\tilde{\delta}\tilde{\delta}\tilde{\delta}\tilde{\delta}\tilde{\delta}\tilde{\delta}\big>_c$, which similarly to the case of the power spectrum, can be further decomposed into SSC and non-SSC parts.
\end{enumerate}
The following subsections address the evaluation of each of these contributions in turn. At the end, we will arrive at a result that is valid for any nonlinear value of the hard modes $k_1, k_1', k_2, k_2'$ and linear values of the soft modes $s_1, s_2$.

\subsection{The $PPP$ term}\label{sec:cov_BB_PPP}

The $PPP$ contribution to the 6-point function of the windowed density contrast is given by
\bq\label{eq:PPP_1}
&&\Big< \tilde{\delta}_W(\vp) \tilde{\delta}_W(\vq) \Big>\Big< \tilde{\delta}_W(\vr) \tilde{\delta}_W(\vp') \Big>\Big< \tilde{\delta}_W(\vq') \tilde{\delta}_W(\vr') \Big> +  {\rm permutations} = \nonumber \\
&&= \Pnl(p)\Pnl(r)\Pnl(q') \tilde{W}(\vp+\vq)\tilde{W}(\vr+\vp')\tilde{W}(\vq'+\vr') +  {\rm permutations},
\eq
where the permutations correspond to all different pairings of six elements in three groups of two. In the second equality, we have implicitly assumed that the window momenta is small compared to the momenta of the bispectrum as  we focus on modes sufficiently inside the survey. Further, the permutation written explicitly above is suppressed by the window function terms: $\tilde{W}(\vq'+\vr') = \tilde{W}(-\vp')$ is small because $p' \gg 1/V_W^{1/3}$; similarly for $\tilde{W}(\vp+\vq)$. The only two permutations that are not suppressed by the window function are
\bq\label{eq:PPP_2}
\Pnl(p)\Pnl(q)\Pnl(r) \Big[\tilde{W}(\vp+\vp')\tilde{W}(\vq+\vq') + \tilde{W}(\vp+\vq')\tilde{W}(\vq+\vp')\Big]\tilde{W}(\vr+\vr').
\eq
One can now replace the 6-point function term in the integrand of Eq.~(\ref{eq:covBB_def}) with the above expression to derive the $PPP$ contribution of the squeezed bispectrum covariance
\bq\label{eq:PPP_3}
\cov^{BB}_{PPP} &\approx& \frac{(2\pi)^9}{V_W^2V_{k_1k_1's_1}V_{k_2k_2's_2}} \int_{k_1}{\rm d}^3\vp\int_{k_1'}{\rm d}^3\vq \int_{s_1}{\rm d}^3\vr \int_{k_2}{\rm d}^3\vp'\int_{k_2'}{\rm d}^3\vq' \int_{s_2}{\rm d}^3\vr' \nonumber \\
&& \delta_D(\vp+\vq+\vr)\delta_D(\vp'+\vq'+\vr') \Big[\delta_D(\vp+\vp') \delta_D(\vq+\vq') + \delta_D(\vp+\vq') \delta_D(\vq+\vp')\Big] \nonumber \\
&& \delta_D(\vr+\vr') \Pnl(p)\Pnl(q)\Pnl(r) \nonumber \\
&=& \frac{(2\pi)^9\delta_{T_1T_2} S_{\rm shape}}{V_W^2V_{k_1k_1's_1}V_{k_2k_2's_2}} \int_{k_1}{\rm d}^3\vp\int_{k_1'}{\rm d}^3\vq \int_{s_1}{\rm d}^3\vr \delta_D(\vp+\vq+\vr) \delta_D(0) \Pnl(p)\Pnl(q)\Pnl(r). \nonumber \\
\eq
In the first equality we have approximated $\tilde{W}(\vk) \approx (2\pi)^3\delta_D(\vk)$, and in the second equality, the integrals over the Dirac deltas yield a constraint that the two triangles $\{k_1, k_1', s_1\}$, $\{k_2, k_2', s_2\}$ must be the same (that is the meaning of the symbol $\delta_{T_1T_2}$; $T$ stands for triangle). The factor $S_{\rm shape}$ is a symmetry factor that is determined by the shape of the triangle: $S_{\rm shape} = 1$ for scalene triangles and $S_{\rm shape} = 2$ for isosceles triangles\footnote{For equilateral triangles, one would have $S_{\rm shape} = 6$, but this is not possible in the squeezed limit.}. When we show numerical results below in Sec.~\ref{sec:results}, we will do so for isosceles triangles. If the bin widths are sufficiently narrow, then the power spectra can be taken out of the integral and we can write
\bq\label{eq:PPP_4}
\cov^{BB}_{PPP} \approx \frac{(2\pi)^6\delta_{T_1T_2} S_{\rm shape}}{V_WV_{k_1k_1's_1}} \Pnl(k_1)\Pnl(k_1')\Pnl(s_1),
\eq
in which we have used $\delta_D(0) = V_W/(2\pi)^3$.

\subsection{The $BB$ term}\label{sec:cov_BB_BB}

The $BB$ contribution to the 6-point function of the windowed density contrast is given by
\bq\label{eq:BB_1}
&&\Big< \tilde{\delta}_W(\vp) \tilde{\delta}_W(\vq) \tilde{\delta}_W(\vr') \Big>_c \Big< \tilde{\delta}_W(\vp') \tilde{\delta}_W(\vq') \tilde{\delta}_W(\vr) \Big>_c + {\rm permutations} \nonumber \\
&=& B_m(\vp, \vq, \vr') B_m(\vp', \vq', \vr) \tilde{W}(\vp+\vq+\vr') \tilde{W}(\vp'+\vq'+\vr) + {\rm permutations},
\eq
where the permutations are all different 10 combinations of six elements into two groups of three. In the equality, similarly to Eq.~(\ref{eq:PPP_1}), we have neglected the dependence of the bispectrum on the window function momenta. 

The contribution of the $BB$ term to Eq.~(\ref{eq:covBB_def}) can thus be written as
\bq\label{eq:BB_2}
\cov^{BB}_{BB} &\approx& \frac{(2\pi)^6}{V_W^2V_{k_1k_1's_1}V_{k_2k_2's_2}} \int_{k_1}{\rm d}^3\vp\int_{k_1'}{\rm d}^3\vq \int_{s_1}{\rm d}^3\vr \int_{k_2}{\rm d}^3\vp'\int_{k_2'}{\rm d}^3\vq' \int_{s_2}{\rm d}^3\vr' \nonumber \\
&& \delta_D(\vp+\vq+\vr)\delta_D(\vp'+\vq'+\vr') \delta_D(\vp+\vq+\vr') \delta_D(\vp'+\vq'+\vr)  \nonumber \\
&& B_m(\vp, \vq, \vr') B_m(\vp', \vq', \vr) \nonumber \\
&+& {\rm permutations},
\eq
where, again, we have approximated $\tilde{W}(\vk) \approx (2\pi)^3\delta_D(\vk)$. One of the above permutations will cancel exactly with the term $B_m(k_1, k_1', s_1)B_m(k_2, k_2', s_2)$ in Eq.~(\ref{eq:covBB_def}), while the remaining 9 permutations (listed explicitly in Appendix \ref{app:BBTPperms}) can all contribute sizeably to the $BB$ term.\footnote{More precisely, for any of the 9 possible permutations, one can always find points in $\{k_1, k_1', s_1\}-\{k_2, k_2', s_2\}$ space in which the permutation is not suppressed by the Dirac delta functions.} In each permutation, one of the four Dirac delta functions is redundant; for the permutation that is explicitly written above, we can replace the four Dirac delta functions with $\delta_D(\vp+\vq+\vr)\delta_D(\vp'+\vq'+\vr')\delta_D(\vr-\vr')\delta_D(0)$ and write
\bq\label{eq:BB_3}
\cov^{BB}_{BB} &=& \frac{(2\pi)^3\delta_{s_1s_2}}{V_WV_{k_1k_1's_1}V_{k_2k_2's_2}} \int_{k_1}{\rm d}^3\vp\int_{k_1'}{\rm d}^3\vq \int_{s_1}{\rm d}^3\vr \int_{k_2}{\rm d}^3\vp'\int_{k_2'}{\rm d}^3\vq' \int_{s_2}{\rm d}^3\vr' \nonumber \\
&& \delta_D(\vp+\vq+\vr)\delta_D(\vp'+\vq'+\vr') \delta_D(\vr-\vr')  B_m(\vp, \vq, \vr') B_m(\vp', \vq', \vr) \nonumber \\
&+& {\rm permutations},
\eq
where the Kronecker delta $\delta_{s_1s_2}$ anticipates that, upon integration, the permutation is only non-vanishing if $s_1$ and $s_2$ are in the same wavenumber bin, and we have used $\delta_D(0) = V_W/(2\pi)^3$. In the squeezed limit, the bispectra in the integrand can be evaluated with the first-order response using Eq.~(\ref{eq:intro1}) (cf.~Sec.~\ref{sec:responses}).

\subsection{The $TP$ term}\label{sec:cov_BB_TP}

The $TP$ contribution to the 6-point function of the windowed density field is given by
\bq\label{eq:TP_1}
&&\Big< \tilde{\delta}_W(\vp) \tilde{\delta}_W(\vr) \Big> \Big< \tilde{\delta}_W(\vp')\tilde{\delta}_W(\vq') \tilde{\delta}_W(\vr') \tilde{\delta}_W(\vq) \Big>_c + {\rm permutations} \nonumber \\
&=& \Pnl(p) T_m(\vp', \vq', \vr', \vq) \tilde{W}(\vp+\vr) \tilde{W}(\vp'+\vq' + \vr'+\vq) + {\rm permutations},
\eq
where the permutations are all 15 different combinations of six elements into a group of four and a group of two. The permutation written explicitly above is suppressed by the window function because $\tilde{W}(\vp+\vr) = \tilde{W}(-\vq)$, which is small for the modes $q \gg 1/V_W^{1/3}$ we consider. Out of the 15 permutations, 6 are suppressed by this reasoning, while 9 can contribute sizeably. As before, when not dealing with the connected 6-point function (cf.~Sec.~\ref{sec:cov_BB_6pt} next), we ignore the dependence of the trispectrum and power spectrum on the window function momenta.

The contribution from this term to Eq.~(\ref{eq:covBB_def}) reads (we keep on using the approximation $\tilde{W}(\vk) \approx (2\pi)^3\delta_D(\vk)$ for modes sufficiently inside the survey.)
\bq\label{eq:TP_2}
\cov^{BB}_{TP} &\approx& \frac{(2\pi)^6}{V_W^2V_{k_1k_1's_1}V_{k_2k_2's_2}} \int_{k_1}{\rm d}^3\vp\int_{k_1'}{\rm d}^3\vq \int_{s_1}{\rm d}^3\vr \int_{k_2}{\rm d}^3\vp'\int_{k_2'}{\rm d}^3\vq' \int_{s_2}{\rm d}^3\vr' \nonumber \\
&& \delta_D(\vp+\vq+\vr)\delta_D(\vp'+\vq'+\vr') \delta_D(\vp+\vp')\delta_D(\vq+\vq'+\vr+\vr') \nonumber \\
&& \Pnl(p) T_m(\vq, \vq', \vr, \vr') + {\rm permutations} \nonumber \\
&=& \frac{(2\pi)^3 \delta_{k_1 k_2}}{V_WV_{k_1k_1's_1}V_{k_2k_2's_2}} \int_{k_1}{\rm d}^3\vp\int_{k_1'}{\rm d}^3\vq \int_{s_1}{\rm d}^3\vr \int_{k_2}{\rm d}^3\vp'\int_{k_2'}{\rm d}^3\vq' \int_{s_2}{\rm d}^3\vr' \nonumber \\
&& \delta_D(\vp+\vq+\vr)\delta_D(\vp'+\vq'+\vr') \delta_D(\vp+\vp') \Pnl(p) T_m(\vq, \vq', \vr, \vr') \nonumber \\
&& + {\rm permutations} 
\eq
where one should now only consider the 9 sizeable permutations; they are all explicitly listed in Appendix \ref{app:BBTPperms}. The second equality above uses the fact that one of the Dirac delta functions is redundant and we can set $\delta_D(0) = V_W/(2\pi)^3$; the Kronecker delta $\delta_{k_1 k_2}$ anticipates that the permutation is only non-vanishing if $k_1$ and $k_2$ are in the same wavenumber bin.

What is left to specify is how to evaluate the trispectra terms in the integrand of Eq.~(\ref{eq:TP_2}). For the case of the permutation that is written explicitly above, the trispectrum corresponds to the squeezed 4-point function with hard modes $\vq, \vq'$ and soft modes $\vr, \vr'$. It can thus be evaluated with responses as (cf.~Sec.~\ref{sec:responses})
\bq\label{eq:TP_3}
&&T_m(\vq, \vq', \vr, \vr') = 2\R_2(q, \mu_{\vq, \vr}, \mu_{\vq, \vr'}, \mu_{\vr, \vr'}, r/r') \Pnl(q)\Plin(r)\Plin(r') \nonumber \\
&+& \R_1(q, \mu_{\vq,\vr+\vr'}) \left[2F_2(-\vr-\vr', \vr)\Plin(r) + 2F_2(-\vr-\vr', \vr')\Plin(r')\right] \Pnl(q) \Plin(|\vr+\vr'|). \nonumber \\
\eq
All remaining permutations (cf.~Appendix \ref{app:BBTPperms}) can be evaluated analogously to the one above, except that for which the trispectrum term is $T_m(\vp, \vq, \vp', \vq') = T_m(\vp, -\vp - \vr, \vp', -\vp' - \vr')$. Interestingly, this particular configuration is the same as that which determines the 4-point function contribution to the matter power spectrum covariance in Eq.~(\ref{eq:NGterm}). The calculation of this specific trispectrum contribution is thus exactly as that described in Sec.~\ref{sec:cov_PP_NG} (cf.~Eq.~(\ref{eq:Tmexp})), with the soft sub-survey triangle mode $\vr$ playing the role of the super-survey mode $\vv$ (note that $\vr = -\vr'$ in this permutation since it contains a term $\delta_D(\vr+\vr')$; cf.~Appendix \ref{app:BBTPperms})\footnote{Note that even though the practical evaluation of the trispectra is the same, in the $TP$ term, the trispectrum depends only on sub-survey modes, and hence, it is not physically rigorous to split and interpret the different contributions into SSC and nonSSC (as it is done in Sec.~\ref{sec:cov_PP_NG}).}.

Overall, Eqs.~(\ref{eq:TP_2}) and (\ref{eq:TP_3}), together with the recipe of Sec.~\ref{sec:cov_PP_NG} for the $T_m(\vp, -\vp + \vr, \vp', -\vp' - \vr)$ permutation, constitute the calculation of our $TP$ term of the covariance matrix of the squeezed matter bispectrum.

\subsection{The connected 6-point function term}\label{sec:cov_BB_6pt}

The connected part of the 6-point function in Eq.~(\ref{eq:covBB_def}) can be worked out as follows
\bq\label{eq:6pt_1}
&&\Big< \tilde{\delta}_W(\vp) \tilde{\delta}_W(\vq) \tilde{\delta}_W(\vr) \tilde{\delta}_W(\vp') \tilde{\delta}_W(\vq') \tilde{\delta}_W(\vr') \Big>_c \nonumber \\
&=& \Bigg[\prod_{a = 1}^6 \int \frac{{\rm d}^3\vv_a}{(2\pi)^3}\tilde{W}(\vv_a)\Bigg] \Big< \tilde{\delta}(\vp-\vv_1) \tilde{\delta}(\vq-\vv_2) \tilde{\delta}(\vr-\vv_3) \tilde{\delta}(\vp'-\vv_4) \tilde{\delta}(\vq'-\vv_5) \tilde{\delta}(\vr'-\vv_6)\Big>_c \nonumber \\
&=& \Bigg[\prod_{a = 1}^6 \int \frac{{\rm d}^3\vv_a}{(2\pi)^3}\tilde{W}(\vv_a)\Bigg] (2\pi)^3 Q_{m,6}(\vp-\vv_1, \vq-\vv_2, \vr-\vv_3, \vp'-\vv_4, \vq'-\vv_5, \vr'-\vv_6) \delta_D(\vv_{123456}) \nonumber \\
&=& \int \frac{{\rm d}^3\vv}{(2\pi)^3} |\tilde{W}(\vv)|^2  Q_{m,6}(\vp, \vq, \vr+\vv, \vp', \vq', \vr'-\vv), \nonumber \\
\eq
where the second equality establishes our definition of the polispectra of the 6-point function $Q_{m,6}$; we have also implicitly used the fact that $\vp+\vq+\vr = \vp'+\vq'+\vr' = 0$, as ensured by the Dirac deltas in Eq.~(\ref{eq:covBB_def}). The third equality uses Eq.~(\ref{eq:useful1}) to simplify the various integrals over the $\vv_a$. Similarly to the connected 4-point function in the case of the power spectrum covariance, here one can also split the contributions to $Q_{m,6}$ into (i) those that vanish for vanishing $\vv$, which is the SSC term $Q_{m,6}^{SSC}$, and (ii) those that remain non-zero in general, which is the remainder of the connected 6-point function contributions $Q_{m, 6}^{nonSSC}$. The above equation can thus be written as
\bq\label{eq:6pt_1}
&&\Big< \tilde{\delta}_W(\vp) \tilde{\delta}_W(\vq) \tilde{\delta}_W(\vr) \tilde{\delta}_W(\vp') \tilde{\delta}_W(\vq') \tilde{\delta}_W(\vr') \Big>_c = \nonumber \\
&& V_W Q_{m, 6}^{nonSSC}(\vp, \vq, \vr, \vp', \vq', \vr') + \int \frac{{\rm d}^3\vv}{(2\pi)^3} |\tilde{W}(\vv)|^2 Q_{m,6}^{SSC}(\vp, \vq, \vr, \vp', \vq', \vr' | \vv),
\eq
where we have ignored the dependence of $Q_{m, 6}^{nonSSC}$ on the window momenta, which is valid for modes deep inside the survey. The bispectrum SSC term $Q_{m,6}^{SSC}$ can be rigorously defined as
\bq\label{eq:6pt_2}
Q_{m,6}^{SSC}(\vp, \vq, \vr, \vp', \vq', \vr' | \vv) = \left[ \lim_{v \to 0} \frac{\partial}{\partial [\Plin(v)]} Q_{m,6}\big(\vp, \vq, \vr+\vv, \vp', \vq', \vr' -\vv \big)\right] \Plin(v).
\eq
A systematic way to derive this term is to draw all of the tree-level diagrams that contribute to the 6-point function $Q_{m,6}$ and keep the permutations that do not vanish if $\vv = 0$; in practice, this corresponds to all of the diagrams with lines that propagate $\vv$. This derivation is outlined with detail in Appendix \ref{app:SSC_BB}; the final result is given by
\bq\label{eq:6pt_3}
Q_{m,6}^{SSC}(\vp, \vq, \vr, \vp', \vq', \vr' | \vv) &=& \raisebox{-0.0cm}{\includegraphicsbox[scale=0.85]{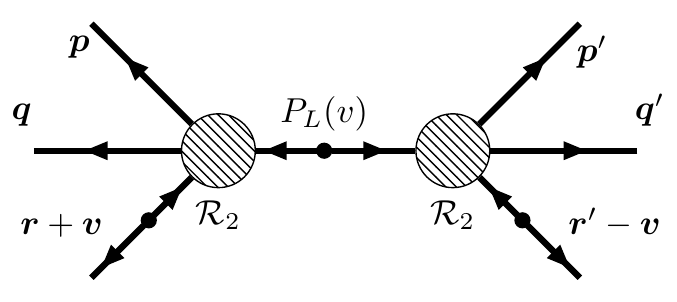}} \nonumber \\
&=& \R_2(p, \mu_{\vp,\vr}, \mu_{\vp,\vv}, \mu_{\vr,\vv}, r/v) \R_2(p', \mu_{\vp',\vr'}, -\mu_{\vp',\vv}, -\mu_{\vr',\vv}, r'/v) \nonumber \\
&\times& \Pnl(p)\Pnl(p')\Plin(r)\Plin(r')\Plin(v).
\eq
Note that the factors of $2!$ that appear in Eq.~(\ref{eq:sqnpt}) do not appear in this equation since there are no soft mode permutations (we have also used $|\vr+\vv|\approx r$). The above equation describes the correlation of the bispectra $B_m(\vp, \vq, \vr)$ and $B_m(\vp', \vq', \vr')$ that is induced by their first-order response to shared super-survey modes $\vv$. Interestingly, for the squeezed configurations that we consider $r \ll p,q$, $r' \ll p',q'$, the first-order bispectrum response \cite{2016PhRvD..94h3528A} turns into the second-order power spectrum response\footnote{Another equivalent point of view is to observe that the squeezed bispectrum is given in terms of the first-order response $\R_1$ (case $n=1$ in Eq.~(\ref{eq:sqnpt})), and hence, the response of the first-order response is a second order effect, i.e., given by $\R_2$.} $\R_2$. Ultimately, this is the reason why we are able to compute the covariance of the squeezed bispectrum using only power spectrum responses, without forcibly requiring bispectrum response functions. 

Reference~\cite{2018PhRvD..97d3532C} has also recently calculated the SSC contribution to the matter bispectrum covariance matrix. This was done both in the context of SPT and the halo model formalism, and the calculation holds for bispectrum configurations beyond the squeezed limit. The SPT derivation of Ref.~\cite{2018PhRvD..97d3532C} should capture the same terms as our calculation if the second-order response $\R_2$ is evaluated at tree-level \cite{responses1}, i.e., not using separate universe simulation measurements. In the nonlinear regime of the hard modes, the halo model calculation of Ref.~\cite{2018PhRvD..97d3532C} is shown to be in good agreement with numerical estimates of the super-sample effect using simulation sub-volumes; there are some relatively small discrepancies between the halo model and the simulations, but which are likely due to the limited accuracy of the halo model. Our calculation is, on the other hand, expected to be accurate for any nonlinear value of the hard modes (but linear soft modes), provided the response coefficients (cf.~Fig.~\ref{fig:Ro}) are measured with separate universe simulations\footnote{The halo model calculation of Ref.~\cite{2018PhRvD..97d3532C} also only includes the response to isotropic density fluctuations, i.e., it does not encompass the response coefficients associated with tidal fields. The contribution from the latter may not be important for the case of angle-averaged three-dimensional bispectra estimates, but could become relevant for the case of galaxy bispectra \cite{2018arXiv180302132S, 2018JCAP...02..022L, akitsu/takada/li} (which is anisotropic due to redshift space distortions) or the bispectrum of projected quantities such as weak lensing shear \cite{completessc}. Our calculation based on the response coefficients of Eq.~(\ref{eq:Ro}) naturally incorporates all of these tidal super-sample effects in the nonlinear regime.}.

Putting it together, the contribution of the connected 6-point function to the covariance of the squeezed bispectrum can be written as
\bq\label{eq:6pt_4}
\cov^{BB}_{6pt} = \cov^{BB}_{6pt, nonSSC} + \cov^{BB}_{6pt, SSC},
\eq
with
\bq
\label{eq:6pt_5_noSS}\cov^{BB}_{6pt, nonSSC} &=& \frac{1}{V_WV_{k_1k_1's_1}V_{k_2k_2's_2}} \int_{k_1}{\rm d}^3\vp\int_{k_1'}{\rm d}^3\vq \int_{s_1}{\rm d}^3\vr \int_{k_2}{\rm d}^3\vp'\int_{k_2'}{\rm d}^3\vq' \int_{s_2}{\rm d}^3\vr' \nonumber \\
&& \delta_D(\vp+\vq+\vr)\delta_D(\vp'+\vq'+\vr')  Q_{m, 6}^{non-SSC}(\vp, \vq, \vr, \vp', \vq', \vr') \\
\label{eq:6pt_5}\cov^{BB}_{6pt, SSC} &=& \frac{1}{V_W^2V_{k_1k_1's_1}V_{k_2k_2's_2}} \int_{k_1}{\rm d}^3\vp\int_{k_1'}{\rm d}^3\vq \int_{s_1}{\rm d}^3\vr \int_{k_2}{\rm d}^3\vp'\int_{k_2'}{\rm d}^3\vq' \int_{s_2}{\rm d}^3\vr' \nonumber \\
&& \delta_D(\vp+\vq+\vr)\delta_D(\vp'+\vq'+\vr') \int \frac{{\rm d}^3\vv}{(2\pi)^3} |\tilde{W}(\vv)|^2 Q_{m, 6}^{SSC}(\vp, \vq, \vr, \vp', \vq', \vr' | \vv), \nonumber \\
\eq
and where $Q_{m, 6}^{SSC}$ is given by Eq.~(\ref{eq:6pt_3}). In this paper, we do not evaluate the non-SSC part of the connected 6-point function $Q_{m, 6}^{nonSSC}$. This term could be calculated using SPT, but the result would only be valid in the perturbative regime. The response approach can be used to resum some contributions to $Q_{m, 6}^{nonSSC}$ by using power spectrum responses $\R_n$ up to $n=4$, as well as general bispectrum response functions \cite{2016PhRvD..94h3528A, 2018PhRvD..97d3532C}; these are, however, calculations that we do not undergo in this paper. We will use the results shown below in Sec.~\ref{sec:results}, however, to argue that the contribution from $Q_{m, 6}^{nonSSC}$ is a negligible one for squeezed configurations.

\section{The matter power spectrum-squeezed bispectrum cross-covariance}\label{sec:cov_BP}

Joint analyses of the power spectrum and bispectrum require also the knowledge of the corresponding cross-covariance. The steps of the derivation are very similar to those taken in the last section for the bispectrum, and so below we skip repeating analogous details.

\subsection{The decomposition of the power spectrum and bispectrum cross-covariance}\label{sec:cov_BP_estimator}

The cross-covariance between the power spectrum and bispectrum estimators of Eqs.~(\ref{eq:Pk3D}) and (\ref{eq:sqB_estimator}), respectively, is given by
\bq\label{eq:covBP_def}
\cov^{BP} &\equiv& \cov^{BP}(k_1, k_1', s_1, k_2) \nonumber \\
&=& \Big<\hat{B}_W(k_1, k_1', s_1)\hat{P}_W(k_2)\Big> - \Big<\hat{B}_W(k_1, k_1', s_1)\Big>\Big<\hat{P}_W(k_2)\Big> \nonumber \\
&=& \frac{1}{V_W^2V_{k_1k_1's_1}V_{k_2}} \int_{k_1}{\rm d}^3\vp\int_{k_1'}{\rm d}^3\vq \int_{s_1}{\rm d}^3\vr \int_{k_2}{\rm d}^3\vp' \delta_D(\vp+\vq+\vr) \nonumber \\
&&\Big< \tilde{\delta}_W(\vp) \tilde{\delta}_W(\vq) \tilde{\delta}_W(\vr) \tilde{\delta}_W(\vp') \tilde{\delta}_W(-\vp')  \Big>_c \nonumber \\
&-& B_m(k_1, k_1', s_1)P_m(k_2).
\eq
The 5-point function can be split into two distinct types of contributions: (i) one given by the product of 2- and 3-point functions, which we call the $BP$ term; and (ii) one determined by the connected 5-point correlation function, which can be split into SSC and non-SSC terms.

\subsection{The $BP$ term}\label{sec:cov_BP_BP}

The $BP$ contribution to the 5-point function of the windowed density contrast is given by
\bq\label{eq:BP_1}
&&\Big< \tilde{\delta}_W(\vp') \tilde{\delta}_W(\vq) \tilde{\delta}_W(\vr) \Big>_c \Big< \tilde{\delta}_W(\vp) \tilde{\delta}_W(-\vp') \Big> + {\rm permutations} \nonumber \\
&=& B_m(\vp', \vq, \vr) P_m(p) \tilde{W}(\vp'+\vq+\vr) \tilde{W}(\vp-\vp') + {\rm permutations},
\eq
and there are a total of six sizeable permutations (in addition to one which cancels exactly with the term $B_m(k_1, k_1', s_1)P_m(k_2)$ term in Eq.~(\ref{eq:covBP_def})). The $BP$ cross-covariance contribution then follows as
\bq\label{eq:BP_2}
\cov^{BP}_{BP} &\approx& \frac{(2\pi)^3\delta_{k_1k_2}}{V_WV_{k_1k_1's_1}V_{k_2}} \int_{k_1}{\rm d}^3\vp\int_{k_1'}{\rm d}^3\vq \int_{s_1}{\rm d}^3\vr \int_{k_2}{\rm d}^3\vp' \delta_D(\vp+\vq+\vr)\delta_D(\vp - \vp') \nonumber \\
&& B_m(\vp', \vq, \vr) P_m(p) + {\rm permutations},
\eq
where we have approximated $\tilde{W}(\vk) \approx (2\pi)^3\delta_D(\vk)$ and used $\delta_D(0) = V_W/(2\pi)^3$. All of the above permutations are written explicitly in Appendix \ref{app:BBTPperms} and the squeezed bispectrum can be evaluated with the $\R_1$ response using Eq.~(\ref{eq:intro1}) (cf.~Sec.~\ref{sec:responses}).

\subsection{The connected 5-point function term}\label{sec:cov_BP_5pt}

The connected part of the 5-point function in Eq.~(\ref{eq:covBP_def}) can be worked out as follows
\bq\label{eq:5pt_1}
&&\Big< \tilde{\delta}_W(\vp) \tilde{\delta}_W(\vq) \tilde{\delta}_W(\vr) \tilde{\delta}_W(\vp') \tilde{\delta}_W(-\vp')  \Big>_c \nonumber \\
&=& \Bigg[\prod_{a = 1}^5 \int \frac{{\rm d}^3\vv_a}{(2\pi)^3}\tilde{W}(\vv_a)\Bigg] \Big< \tilde{\delta}(\vp-\vv_1) \tilde{\delta}(\vq-\vv_2) \tilde{\delta}(\vr-\vv_3) \tilde{\delta}(\vp'-\vv_4) \tilde{\delta}(-\vp'-\vv_5) \Big>_c \nonumber \\
&=& \Bigg[\prod_{a = 1}^5 \int \frac{{\rm d}^3\vv_a}{(2\pi)^3}\tilde{W}(\vv_a)\Bigg] (2\pi)^3 Q_{m,5}(\vp-\vv_1, \vq-\vv_2, \vr-\vv_3, \vp'-\vv_4, -\vp'-\vv_5) \delta_D(\vv_{12345}) \nonumber \\
&=& \int \frac{{\rm d}^3\vv}{(2\pi)^3} |\tilde{W}(\vv)|^2  Q_{m,5}(\vp, \vq, \vr+\vv, \vp', -\vp' - \vv), \nonumber \\
\eq
where the polispectrum associated with the 5-point function $Q_{m, 5}$ is defined by the second equality above. Similarly to the case of the connected 4- and 6-point functions discussed in the previous sections, the connected 5-point function is also decomposed into SSC and non-SSC contributions, $Q_{m, 5}= Q_{m, 5}^{nonSSC}  + Q_{m, 5}^{SSC}$. The SSC contribution is given by (the derivation steps are analogous to those taken in Appendix \ref{app:SSC_BB} for $Q_{m,6}^{SSC}$)
\bq\label{eq:5pt_2}
Q_{m,5}^{SSC}(\vp, \vq, \vr, \vp', -\vp' | \vv) &=& \raisebox{-0.0cm}{\includegraphicsbox[scale=0.85]{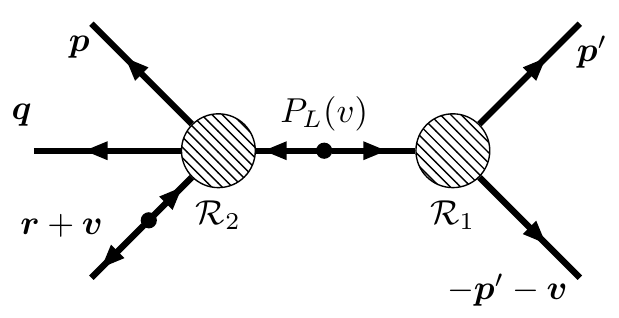}} \nonumber \\
&=&  \R_2(p, \mu_{\vp,\vr}, \mu_{\vp,\vv}, \mu_{\vr,\vv}, r/v) \R_1(p', -\mu_{\vp',\vv}) \nonumber \\
&\times& \Pnl(p)\Pnl(p')\Plin(r)\Plin(v).
\eq
The contribution from the connected part can thus be written as
\bq\label{eq:5pt_3}
\cov^{BP}_{5pt} = \cov^{BP}_{5pt, nonSSC} + \cov^{BP}_{5pt, SSC},
\eq
with
\bq\label{eq:5pt_4}
\cov^{BP}_{5pt, nonSSC} &=& \frac{1}{V_WV_{k_1k_1's_1}V_{k_2}} \int_{k_1}{\rm d}^3\vp\int_{k_1'}{\rm d}^3\vq \int_{s_1}{\rm d}^3\vr \int_{k_2}{\rm d}^3\vp' \nonumber \\
&& \delta_D(\vp+\vq+\vr)  Q_{m, 5}^{nonSSC}(\vp, \vq, \vr, \vp', \vp') \\
\cov^{BP}_{5pt, SSC} &=& \frac{1}{V_W^2V_{k_1k_1's_1}V_{k_2}} \int_{k_1}{\rm d}^3\vp\int_{k_1'}{\rm d}^3\vq \int_{s_1}{\rm d}^3\vr \int_{k_2}{\rm d}^3\vp' \nonumber \\
&& \delta_D(\vp+\vq+\vr) \int \frac{{\rm d}^3\vv}{(2\pi)^3} |\tilde{W}(\vv)|^2 Q_{m, 5}^{SSC}(\vp, \vq, \vr, \vp', -\vp' | \vv). \nonumber \\
\eq
In this paper, we do not carry out the calculation of the non-SSC part of the connected 5-point function. Similarly to the case of the non-SSC part of the connected 6-point function, part of the contribution can nonetheless be captured in the nonlinear regime using higher-order power spectrum responses and bispectrum responses.

\section{Quantitative results}\label{sec:results}

In this section, we display a few numerical results of the equations derived in the last sections. We consider a spherical survey with $V_W = 50\ {\rm Gpc^3} / h^3$, for which the Fourier transform of the window function is given by
\bq\label{eq:swf}
|\tilde{W}(\vv)|^2 = \left[\frac{3j_1(vR_W)}{vR_W} V_W\right]^2,
\eq
where $R_W = (3V_W/(4\pi))^{1/3}$ and $j_1$ is the first-order spherical Bessel function. For simplicity, we focus on isosceles configurations of the squeezed bispectrum $B_m(k_1, k_1', s_1)$, with $k_1' = k_1$. We use $30$ wavenumber bins equally spaced in log-scale from $5/R_W = 0.002\ h/{\rm Mpc}$ to $2\ h/{\rm Mpc}$. We consider triangles to be squeezed if $k_1 > 5s_1$: the corrections to the equations derived in the previous sections scale as $(s_1/k_1)^2$ (cf.~Eq.~(\ref{eq:Rndef})), and hence, this choice ensures that the corrections are kept below $\lesssim 5\%$ for the least squeezed triangles. This is also why the minimum mode we consider is $5/R_W$ as it ensures that super-survey modes $v \lesssim 1/R_W$ are sufficiently soft compared to the sub-survey modes. We also have $s_1 < 0.05\ h/{\rm Mpc}$ to ensure that the soft sub-survey modes are in the linear regime of structure formation. In total, this yields 217 isosceles squeezed bispectrum configurations.

The covariance matrix of the angle-averaged squeezed bispectrum in isosceles configurations depends on 4 variables, $k_1, k_2, s_1, s_2$. To facilitate displaying the results, we show the covariance as a function of the triangles $\{k_1, k_1, s_1\}$ and $\{k_2, k_2, s_2\}$, which we rank order by increasing hard-mode, and for fixed hard-mode, by increasing soft-mode. Explicitly, labeling a triangle as $\{i, i, j\}$ with $i$ the bin of the hard modes and $j$ the bin of the soft mode, our ordered list of triangles is
\bq\label{eq:trianglelist}
\{8,8,0\}, \{9, 9, 0\}, \{9, 9, 1\}, \{10, 10, 0\}, \{10,10,1\}, \{10, 10,2\}, \cdots , \{29,29,0\}, \cdots, \{29,29,13\}. \nonumber \\
\eq
The 8th bin is the first that is 5 times the 0th bin, which is why the first triangle is $\{8,8,0\}$. We adopt the following cosmological parameters in a flat $\Lambda$CDM model: total matter density $\Omega_m = 0.27$, baryonic matter density $\Omega_b = 0.0469$, dimensionless Hubble parameter $h = 0.70$, scalar spectral index $n_s = 0.95$ and r.m.s.~of the matter fluctuations today $\sigma_8 = 0.8$. All our results are for redshift $z = 0$. We evaluate power spectra using {\sc CAMB} \cite{camb} with the {\sc HALOFIT} \cite{2003MNRAS.341.1311S} implementation of Ref.~\cite{2012ApJ...761..152T}, the response functions are evaluated using Eq.~(\ref{eq:Ro}), and the various bin and angle averages are carried out with Monte Carlo integration (cf.~Appendix \ref{app:numericalevaluation}).

\subsection{Matter bispectrum covariance results}\label{sec:results_bb}

\begin{figure}[t!]
        \centering
        \includegraphics[width=\textwidth]{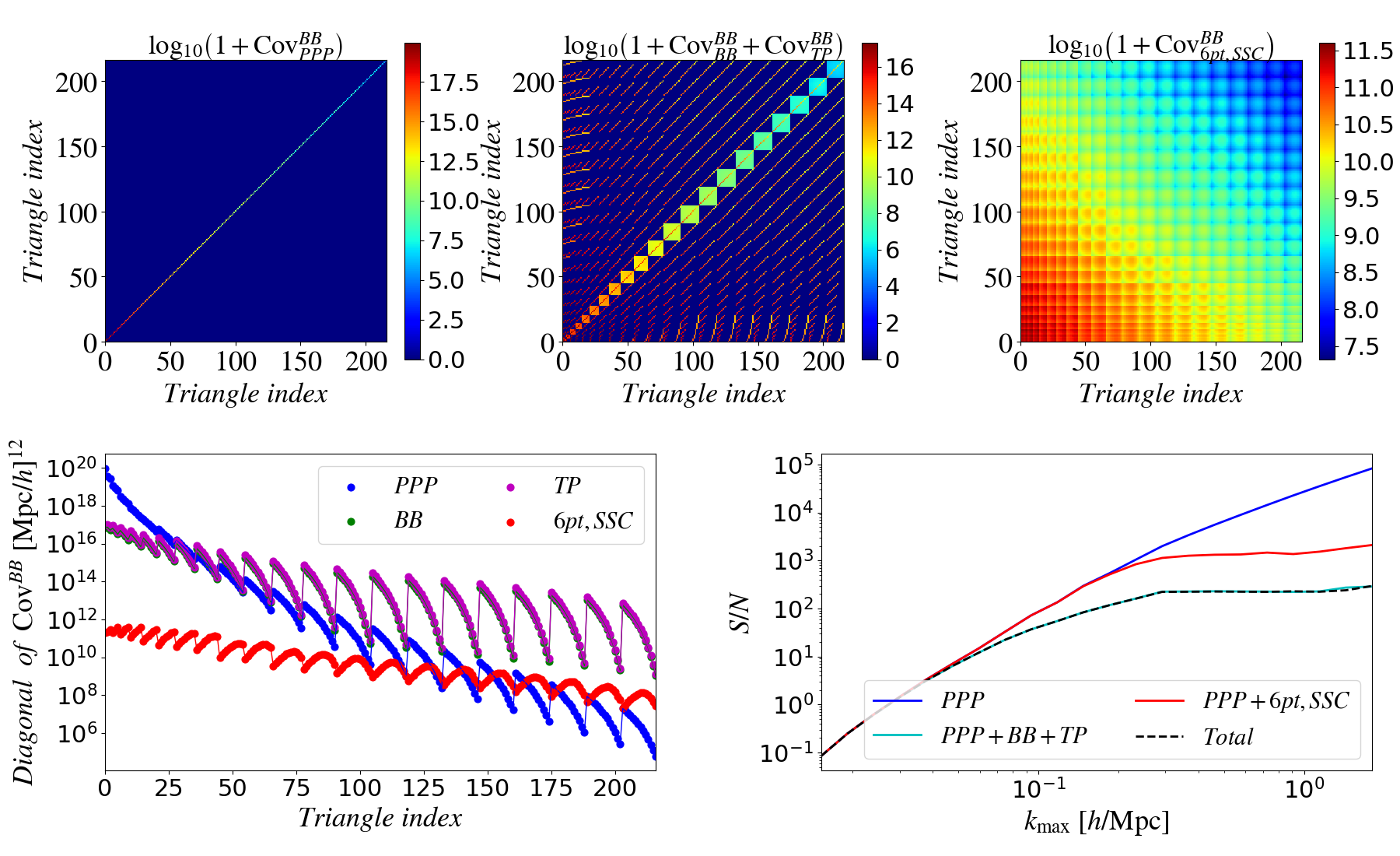}
        \caption{The upper panels show the $PPP$ (left), $BB+TP$ (middle) and $6pt,SSC$ (right) contributions to the covariance matrix of the squeezed matter bispectrum. The indexing of the triangles is that of Eq.~(\ref{eq:trianglelist}). We show the logarithm of $1$ plus the actual covariance contribution to highlight better their structure (note that the color scale is also different in the three panels); in particular, in the upper left and upper right panels, the dark blue color indicates vanishing contribution. The lower left panel shows the diagonal (i.e., same triangle) of the various terms, as labeled (the $BB$ and $TP$ cases are barely distinguishable on a log-scale). The lower right panel shows the cumulative signal-to-noise ratio (defined in Eq.~(\ref{eq:sn}); the black and cyan curves are nearly overlapping).}
\label{fig:bb}
\end{figure}

The upper panels of Fig.~\ref{fig:bb} show the $\cov^{BB}_{PPP}$, $\cov^{BB}_{BB} + \cov^{BB}_{TP}$ and $\cov^{BB}_{6pt, SSC}$ contributions, as labeled. The $6pt,SSC$ term is non-zero for all triangles, but the $PPP$, $BB$ and $TP$ terms are not. The $PPP$ term is only non-zero if the two triangles are the same ($k_1=k_2, s_1=s_2$; this is the diagonal of the matrix). The $BB$ and $TP$ terms are non-zero only if the two hard modes are the same ($k_1=k_2$; non-zero blocks along the diagonal), if the hard mode in one triangle is equal to the soft mode in the other ($k_1=s_2$ or $k_2 = s_1$; nearly vertical stripes near the plot axes), or if the two soft modes are the same ($s_1 = s_2$; remainder of the non-zero terms). The lower left panel shows these contributions along the diagonal and permits a better visualization of their relative size. The $PPP$ term dominates for the lowest triangle indices, which corresponds to the largest distance scales and is as expected. When the hard mode of the triangles approach nonlinear scales, $k_1 = k_2 \approx 0.1 h/{\rm Mpc}$ (approximately index $30$), the $BB$ and $TP$ terms begin to dominate, and continue to do so for all smaller-scale triangles (higher triangle index). These two terms have the same order of magnitude as their calculation involves the same number (four) of power spectra terms\footnote{The 1-loop tripsectrum in one of the $TP$ permutations actually depends on five power spectra, but it involves also an integral over the loop momenta.}. Interestingly, the super-sample contribution to the squeezed-matter bispectrum remains subdominant for all scales probed.

It is instructive to understand the origin of the size hierarchy of the $\cov^{BB}_{PPP}$, $\cov^{BB}_{BB}$, $\cov^{BB}_{TP}$ and $\cov^{BB}_{6pt, SSC}$ contributions shown in Fig.~\ref{fig:bb}. A crude (but sufficient to the purpose) order of magnitude estimate of the relative size of each term can be obtained from the equations derived in the previous sections by setting the response functions to unity (cf.~Fig.~\ref{fig:Ro}) and approximating the power spectra as constant inside each wavenumber bin. Doing so, the $\cov^{BB}_{PPP}$, $\cov^{BB}_{BB}$ and $\cov^{BB}_{6pt, SSC}$ terms can be shown to have the following dependencies along the diagonal (dropping also a few numerical pre-factors)
\bq
\label{eq:res_ppp_simple} \cov^{BB}_{PPP}  &\sim& \frac{\Pnl(k_1)^2\Plin(s_1)}{V_W} \frac{1}{(k_1\Delta{k_1})^2s_1\Delta{s_1}} ,\\ 
\label{eq:res_bb_simple}  \cov^{BB}_{BB}  &\sim&  \frac{\Pnl(k_1)^2\Plin(s_1)^2}{V_W} \left[\frac{1}{s_1^2 \Delta{s_1}} + \frac{1}{k_1^2 \Delta{k_1}}\right] \approx \frac{\Pnl(k_1)^2\Plin(s_1)^2}{V_W} \frac{1}{s_1^2 \Delta{s_1}}, \\
\label{eq:res_ssc_simple} \cov^{BB}_{6pt, SSC} &\sim& \frac{\Pnl(k_1)^2\Plin(s_1)^2}{V_W^2} \int \frac{{\rm d}\vv^3}{(2\pi)^3}|\tilde{W}(\vv)|^2\Plin(v),
\eq
where $\Delta{k}$ denotes the width of the bin of the mode $k$ and, in the $BB$ equation, the second equality neglects the contribution from the second term inside the square brackets as it is subdominant in the squeezed limit; further, the dependencies of the $\cov^{BB}_{TP}$ term are similar to those of the $BB$ term, so we skip writing them explicitly. The explicit dependencies on the amplitude of the bispectrum modes comes from the Fourier integration volume factors. These equations readily explain the hierarchy observed in the lower left panel of Fig.~\ref{fig:bb}. For instance, the ratio of the $PPP$ to the $BB$ term scales as ${\cov^{BB}_{PPP}}/{\cov^{BB}_{BB}} \sim {s_1}/{((k_1\Delta k_1)^2\Plin(s_1))}$, which leaves apparent how the $PPP$ term can become smaller with increasing $k_1$ (i.e., increasing triangle index)\footnote{Actually, the value of $s_1/\Plin(s_1)$ can also increase with the triangle index (cf.~Eq.~(\ref{eq:trianglelist})), but $k_1^2\Delta k_1^2$ does so faster.}, as shown in the lower left panel of Fig.~\ref{fig:bb}. On the other hand, the ratio of the $6pt, SSC$ to the $BB$ contribution can be estimated as
\bq\label{eq:res_ratio_simple}
\frac{\cov^{BB}_{6pt, SSC}}{\cov^{BB}_{BB}} &\sim& s_1^2 \Delta s_1 \frac{1}{V_W} \int \frac{{\rm d}\vv^3}{(2\pi)^3}|\tilde{W}(\vv)|^2\Plin(v) \nonumber \\
&\approx& 10^5 \times s_1^2 \Delta s_1,
\eq
where the second equality uses our assumed survey geometry and volume. This equation shows that the $6pt,SSC$ contribution can become negligible compared to $BB$ (and $TP$) if the amplitude of the soft mode $s_1$ is sufficiently small. Indeed, in our calculation, the soft mode must be kept inside the linear regime of structure formation $s_1 \ll \knl$, which pinpoints the origin behind the smallness of the $6pt,SSC$ term observed in Fig.~\ref{fig:bb}. Naturally, the relative size of the $6pt,SSC$ and $BB$ terms depends on the survey volume via the integral over the window function. We have checked explicitly, however, that if $s_1$ is in the linear regime (and $\Delta s_1$ is not abusively large) then the $6pt, SSC$ contribution is always subdominant at least for $V_W \gtrsim 5\ {\rm Gpc}^3/h^3$.

The lower right panel of Fig.~\ref{fig:bb} shows the cumulative signal-to-noise ratio, which is defined as
\bq\label{eq:sn}
\left(\frac{S}{N}\right)^2_{<k_{\rm max}} = \sum_{\text{All triangles with}\atop k_1,k_2 < k_{\rm max}}  B_m(k_1, k_1, s_1) [\cov^{BB}(k_1, s_1, k_2, s_2)]^{-1} B_m(k_2, k_2, s_2),
\eq
where the sum runs over all triangles with hard mode smaller than $k_{\rm max}$. The result is shown for varying subsets of covariance contributions and is in line with the relative size of the various terms discussed above. In  particular, the $BB$ and $TP$ terms dominate the degradation in the signal-to-noise relative to the $PPP$ case, with the $6pt,SSC$ term accounting only for negligible degradation (cf.~nearly indistinguishable cyan and black curves in the lower right panel of Fig.~\ref{fig:bb}).

A take away message is therefore that neglecting the contribution from the $6pt,SSC$ term in real data (or forecast) applications of the squeezed bispectrum is likely to be a good approximation, provided the soft mode of the triangle is in the linear regime of structure formation. In fact, this conclusion on the smallness of the $6pt, SSC$ term can be extended to the rest of the connected 6-point function $\cov^{BB}_{6pt, nonSSC}$, given that both share the same scalings (or lack thereof) with momenta amplitudes and are given by connected 6-point function terms of similar magnitude (cf.~the derivation of the $6pt, SSC$ term in Appendix \ref{app:SSC_BB}, which displays also the $6pt,nonSSC$ terms). The $\cov^{BB}_{6pt, nonSSC}$ was the only term that we did not evaluate with responses (cf.~Sec.~\ref{sec:cov_BB_6pt}), and hence, its negligible contribution effectively makes our calculation of the squeezed  bispectrum covariance complete\footnote{As a (very crude) way to estimate the impact of the missing $6pt,nonSSC$ term, we have explicitly checked that our total signal-to-noise results do not change if we double the size of the $6pt,SSC$ term.}. 

It is also interesting to interpret part of the analysis done in Ref.~\cite{2018arXiv180206762D} in light of our results above. In Ref.~\cite{2018arXiv180206762D}, the authors studied the forecast constraining power on local primordial non-Gaussianity  of a number of large-scale structure statistics, including the squeezed matter bispectrum. There, the latter is modeled in terms of position-dependent power spectra \cite{2014JCAP...05..048C}, which is effectively the same as using the first-order power spectrum response function $\R_1$, as done in this paper (cf.~Eq.~(\ref{eq:intro1})). The covariance calculated in Ref.~\cite{2018arXiv180206762D} includes the $PPP$ term and the permutations of the $BB$ and $TP$ terms that are non-zero if the soft modes of the triangles are in the same bin (these are the $A$-type permutations in Appendix \ref{app:BBTPperms}). We have explicitly checked that these $BB$ and $TP$ permutations are indeed the dominant ones: the lower panels of Fig.~\ref{fig:bb} remain virtually the same if we set all remaining permutations to zero (these are also the permutations that give the right-hand side of Eq.~(\ref{eq:res_bb_simple})). Reference \cite{2018arXiv180206762D} also does not include the connected 6-point function contribution, but we have argued above that one is justified to neglect it if the soft squeezed bispectrum mode is small. Hence, despite the varying level of completeness, the covariance matrices in the two works should capture all of the dominant contributions\footnote{An exact comparison is hard to establish because of the different notations and exact formalism. Our expressions are also valid deep in the linear regime of the hard bispectrum modes as we use response measurements from separate universe simulations.} and should thus effectively lead to the same conclusions when used in practical applications. For instance, and importantly, both works underline the necessity to take into account contributions beyond the $PPP$ in real data and forecast studies using the squeezed bispectrum.

We also stress that the conclusion we draw here on the negligible contribution of the $SSC$ term is associated with the squeezed limit, and may not necessarily hold for more general configurations of the bispectrum. In fact, Ref.~\cite{2018PhRvD..97d3532C} shows that the $SSC$ term can contribute with $\approx 30\%$ at $k \approx 0.5\ h/{\rm Mpc}$ for equilateral configurations of the matter bispectrum ($z = 0$; see Fig.~5 there); further, Ref.~\cite{2013MNRAS.429..344K} also shows that the halo-sample variance contribution (which is part of the $SSC$ term) to the weak lensing bispectrum can dominate the total covariance for multipoles $\ell > 10^3$ in equilateral configurations (see also Ref.~\cite{2018arXiv181207437R}).

\subsection{Matter power spectrum and squeezed bispectrum cross-covariance results}\label{sec:results_joint}

\begin{figure}[t!]
        \centering
        \includegraphics[width=\textwidth]{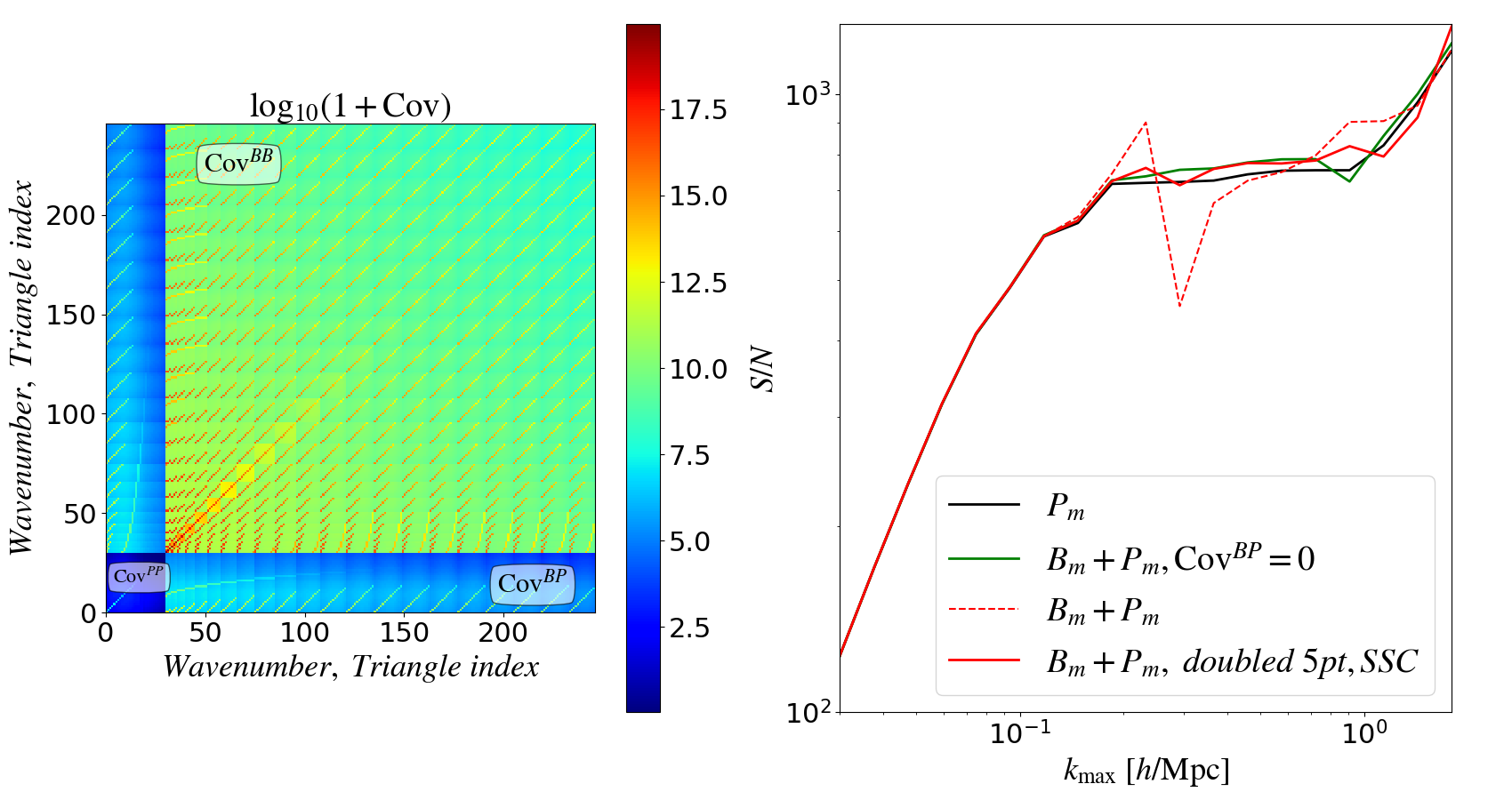}
        \caption{The left panel shows the total covariance matrix for a joint power spectrum-squeezed limit bispectrum {\it observable} vector. The square in the lower right corner is the power spectrum covariance ($\cov^{PP}$), the big square in the upper right corner is the bispectrum covariance ($\cov^{BB}$) and the two rectangular bands near the axes are the cross-covariance term ($\cov^{BP}$), as labeled. Note that these different contributions have different dimensions, hence their marked amplitude difference ($\cov^{PP}, \cov^{BB}, \cov^{BP}$ have dimensions of $L^6, L^{12}, L^9$, respectively, where $L$ is a unit length). The right panel shows the cumulative signal-to-noise for the power spectrum, and joint power spectrum-bispectrum, as labeled. The dashed red line corresponds to using all of the contributions derived in this paper and the green line sets $\cov^{BP}$ to zero, i.e., it treats the power spectrum and bispectrum as independent. The solid red line shows the same as the dashed red line, but with the $5pt,SSC$ contribution doubled.}
\label{fig:total}
\end{figure}

The left panel of Fig.~\ref{fig:total} shows the total covariance matrix for a joint power spectrum and squeezed bispectrum {\it observable} vector (the vector first contains the power spectrum and then the bispectrum with the ordering of Eq.~(\ref{eq:trianglelist})). In the cross-covariance result (rectangular bands near the axes), the $\cov^{BP}_{BP}$ contribution is only non-vanishing if one of the triangle sides is equal to the power spectrum momentum, i.e., $s_1 = k_2$ (brighter stripes near the plot axis) or $k_1 = k_2$. The $\cov^{BP}_{5pt,SSC}$ is non-vanishing for general $\{k_1, k_1, s_1\}$, $k_2$. By taking the same simplifying steps that led to Eqs.~(\ref{eq:res_ppp_simple})-(\ref{eq:res_ssc_simple}), we can estimate the size of the $BP$ and $5pt,SSC$ contributions as
\bq
\label{eq:res_bp_simple}         \cov^{BP}_{BP}  &\sim&  \frac{\Pnl(k_1)\Plin(s_1)\Pnl(k_2)}{V_W} \left[\frac{\delta_{s_1k_2}}{s_1^2 \Delta{s_1}} + \frac{\delta_{k_1k_2}}{k_1^2 \Delta{k_1}}\right] , \\
\label{eq:res_5ptssc_simple} \cov^{BP}_{5pt, SSC} &\sim& \frac{\Pnl(k_1)\Plin(s_1)\Pnl(k_2)}{V_W^2} \int \frac{{\rm d}\vv^3}{(2\pi)^3}|\tilde{W}(\vv)|^2\Plin(v),
\eq
which yields (using our assumed window function and keeping only the $\delta_{s_1k_2}$ part in the $BP$ term because it is larger in the squeezed limit)
\bq\label{eq:bp_ratio}
\left.\frac{\cov^{BP}_{5pt, SSC}}{\cov^{BP}_{BP}}\right\vert_{s_1=k_2} \sim 10^5 \times s_1^2 \Delta s_1.
\eq
This equation displays the same scaling as in Eq.~(\ref{eq:res_ratio_simple}), which we used in the last section to explain the unimportance of the $SSC$ contribution relative to the $BB$ and $TP$ terms. This could motivate us to conclude that, just like for the case of the bispectrum covariance, in the cross-covariance case it is also a good approximation to neglect the contribution from the connected correlators. We will argue next however why this could be a premature conclusion.

The right panel of Fig.~\ref{fig:total} shows, as a function of $k_{\rm max}$, the cumulative signal-to-noise ratio for the power spectrum and joint power spectrum-squeezed bispectrum, as labeled. The signal-to-noise is defined analogously to Eq.~(\ref{eq:sn}), but including now sums over the power spectrum modes as well. The dashed red curve shows the result using all of the covariance terms derived in this paper, which for the cross-covariance part include $\cov^{BP}_{BP}$, $\cov^{BP}_{5pt,SSC}$, but not $\cov^{BP}_{5pt,nonSSC}$ (cf.~Table \ref{table:contsummary}). The strong oscillations exhibited by the dashed red curve are not of physical origin, but instead caused by numerical instabilities associated with inverting an ill-conditioned covariance matrix. A possibility is that these instabilities may arise from the absence of the $\cov^{BP}_{5pt,nonSSC}$ term\footnote{We have checked that progressively increasing the precision of the MC integrals (cf.~Appendix \ref{app:numericalevaluation}) does not seem to alleviate the observed instabilities.}. This term has the same order of magnitude and structure as that of $\cov^{BP}_{5pt,SSC}$, and hence, a (very) crude way to incorporate its contribution to the covariance matrix is to double the size of the $\cov^{BP}_{5pt,SSC}$ term. This is shown by the solid red curve, which is indeed appreciably smoother compared to the dashed curve. A hypothesis is therefore that including the $5pt,nonSSC$ contribution could yield a joint power spectrum-squeezed bispectrum covariance matrix that is more stable under inversion. 

This observation alone does not explain why the connected 5-point function could be important in the case of the cross-covariance, but the connected 6-point function term can be neglected in $\cov^{BB}$, despite both displaying the same relative size compared to the corresponding disconnected pieces (cf.~Eqs.~(\ref{eq:res_ratio_simple}) and (\ref{eq:bp_ratio})). The reason could be associated with the structure of the various covariance matrix contributions. In particular, the dominant $BB$,$TP$ contributions ($A$-type permutations in Appendix \ref{app:BBTPperms}) to the bispectrum covariance span the whole $\{k_1, k_1, s_1\}-\{k_2, k_2, s_2\}$ space (even if sparsely, i.e., they are only non-zero when $s_1=s_2$), and always contribute to the diagonal. On the other hand, the large $BP$ permutations ($H$-type permutations in Appendix \ref{app:BBTPperms}) contribute only to the entries of the cross-covariance where the soft triangle mode is equal to the power spectrum mode, and these entries may not be as important to the inverse of the covariance matrix.

We stress that the above explanations are only tentative ones and that a robust assessment of the relative importance of the various $\cov^{BP}$ contributions and their connection to the behavior in the right-panel of Fig.~\ref{fig:total} should probably involve an explicit calculation of the $5pt,nonSSC$ contribution. Within the response approach, this can be done with a combination of SPT, higher-order power spectrum responses and bispectrum responses. It is also relevant to note that in realistic applications to lensing or galaxy bispectra analyses, the diagonal entries of the covariance matrix will be up-weighted by galaxy shape-noise and shot-noise terms, which is expected to render the inversion of the full covariance matrix less sensitive to inaccuracies in the off-diagonal parts. Methods such as singular value decomposition or matrix regularization/preconditioning \cite{2018arXiv181207437R, 2016JCAP...08..005L} can also be useful in attempts to obtain more accurate inverse covariance matrices. In this paper, we refrain from drawing final conclusions on the relative importance of the various contributions to the squeezed bispectrum and power spectrum cross-covariance, and defer a more detailed investigation for future work.

Finally, it is also interesting to compare the gains in signal-to-noise from the combined power spectrum-bispectrum analysis, relative to using the power spectrum alone. The green curve shows the signal-to-noise ratio of the combined {\it observable} when the power spectrum and bispectrum are treated as independent, i.e., $\cov^{BP} = 0$. This can be used as a {\it proxy} for the signal-to-noise expected from a complete calculation of the cross-covariance term\footnote{Note that treating the power spectrum and bispectrum as independent does not necessarily provide a "largest possible" increase in signal-to-noise. Including $\cov^{BP}$ can increase the signal-to-noise as it incorporates the fact that the small scale modes of the power spectrum and bispectrum {\it couple to the same realization} of super-survey and large-scale sub-survey modes \cite{2006PhRvD..74b3522S}. In other words, setting $\cov^{BP} = 0$ {\it double counts} part of the cosmic variance of the small-scale fluctuations.}. Comparing the green and black curves reveals that the joint power spectrum-bipectrum signal-to-noise is strongly dominated by the power spectrum: the increase in signal-to-noise in the joint case is kept below $5\%$ for all $k_{\rm max}$ shown. The subdominant contribution of the bispectrum to the total signal-to-noise is not totally unexpected considering that we are using only squeezed triangles. As a word of caution, we note that although the signal-to-noise ratio is an effective way to quickly estimate the information contained in some observable, robust and final conclusions on the constraining power of the squeezed bispectrum should be obtained at the level of inferred parameter error bars \cite{2006PhRvD..74b3522S, 2017MNRAS.465.1757G, 2018arXiv181002374C, 2018arXiv180206762D, 2018MNRAS.tmp.2989Y}. For instance, signal-to-noise ratios do not inform on the breaking of parameter degeneracies that the bispectrum can induce (e.g., between linear halo bias $b_1$ and the amplitude of density fluctuations $\sigma_8$ \cite{1994PhRvL..73..215F, 1994ApJ...425..392F, 1997MNRAS.290..651M, 2007PhRvD..76h3004S}). Further, cosmologies with local primordial non-Gaussianity leave specific signatures in the squeezed bispectrum \cite{scoccimarro/etal:2004, liguori/etal:2010, 2010CQGra..27l4011D, 2010AdAst2010E..64V, 2001PhRvD..63f3002K, jeong/komatsu:2009b, 2007PhRvD..76h3004S, 2010JCAP...07..002N, 2011JCAP...04..006B, sefusatti/etal:2012, 2016JCAP...06..014T, 2018arXiv180206762D, 2014JCAP...03..032T}, and hence, the considerations made above for Gaussian initial conditions and for the matter density field (i.e.,~not considering scale-dependent galaxy bias) are not very informative for such cases. A more careful assessment of the constraining power of the squeezed bispectrum, which should include realistic applications to galaxy clustering or weak lensing, as well as forecasts on parameter error bars, is, however, beyond the scope of this paper.

\section{Summary and conclusions}\label{sec:summary}

We have used the response approach to perturbation theory to derive the covariance matrix of the angle-averaged squeezed matter bispectrum, $\cov^{BB} = \cov\left(B_m(k_1, k_1', s_1), B_m(k_2, k_2', s_2)\right)$, $s_i \ll k_i, k_i'$ ($i = 1,2$). A key observation we made is that the covariance of the squeezed bispectrum is dominated by perturbation theory terms that correspond to the coupling of two small-scale modes with one or two large-scale modes. This makes it amenable to be evaluated with first- and second-order power spectrum responses, which can be efficiently measured with separate universe simulations. This effectively ends up resulting in a calculation that is complete and predictive for fully nonlinear values of the small scale bispectrum modes $k_i, k_i'$, with the large-scale modes $s_i$ in the linear regime of structure formation.  

The covariance of the matter bispectrum is determined by a specific configuration of the 6-point matter correlation function (cf.~Eq.~(\ref{eq:covBB_def})). We have organized its derivation by the different types of disconnected and connected terms that contribute to it. Namely, we dubbed by $PPP$ those permutations given by three power spectra, $BB$ those given by two bispectra, $TP$ those given by the trispectrum and the power spectrum, $6pt, SSC$ the super-sample contribution to the connected 6-point function and $6pt,nonSSC$ the rest of the connected 6-point function. We have evaluated all of these terms explicitly, except the $6pt,nonSSC$ one. Table \ref{table:contsummary} summarizes all the terms captured by our calculation.

One of the main conclusions of our numerical results is that the $6pt,SSC$ term contributes only negligibly to the total error budget of analyses using the squeezed bispectrum (cf.~Fig.~\ref{fig:bb}). This can be traced back to the dependencies of the various terms on the size of the bispectrum momenta, which appear in Fourier integration volume factors. More specifically, the dependencies on $s_i, k_i$ cancel out in the $6pt,SSC$ term, but the $BB$ and $TP$ terms have permutations that scale as $s_i^{-2}$ and which dominate as the $s_i$ are in the linear regime of structure formation $s_i \ll \knl$ (cf.~discussion around Eq.~(\ref{eq:res_ratio_simple})).

By the same reasons, the conclusion on the smallness of the $6pt,SSC$ term can be extended to the rest of the connected 6-point function term $6pt,nonSSC$, which cannot be evaluated solely with power spectrum responses. A corollary of this observation is that the calculation presented here, which uses only the power spectrum and its response functions, is sufficient to capture all of the sizeable contributions to the squeezed bispectrum covariance.

Our numerical results also show that the off-diagonal contributions to the covariance matrix yield significant degradation in the signal-to-noise of the bispectrum (this degradation is almost single-handedly due to the $BB$ and $TP$ terms; cf.~Fig.~\ref{fig:bb}). Specifically, considering only the $PPP$ term in the bispectrum covariance (which has been done in some literature for simplicity), significantly overestimates the true cosmological information contained in the bispectrum already at $k \gtrsim 0.1\ h/{\rm Mpc}$. When all covariance terms are taken into account, then the squeezed bispectrum barely contributes to the signal-to-noise of joint power spectrum-bispectrum analysis (cf.~Fig.~\ref{fig:total}). We note, however, that a proper assessment of the constraining power of the bispectrum should be done at the level of parameter constraints, and not simply based on signal-to-noise considerations.

We have also evaluated the power spectrum-bispectrum cross-covariance $\cov^{BP} = \cov\left(B_m(k_1, k_1', s_1), P_m(k_2)\right)$, which is determined by the 5-point matter correlation function. We dubbed by $BP$ the permutations that are given by the bispectrum and power spectrum, by $5pt,SSC$ the super-sample contribution to the connected 5-point function and by $5pt,nonSSC$ the rest of the connected 5-point function. We have evaluated all of these terms, except the $5pt,nonSSC$ one (cf.~Table \ref{table:contsummary}). We pointed out that despite the $BP$ term being larger in size in covariance entries where all terms contribute (cf.~Eq.~(\ref{eq:bp_ratio})), the structure of the cross-covariance matrix may actually imply that the connected 5-point terms cannot be ignored if the matrix is to be stable under inversion (cf.~Sec.~\ref{sec:results_joint}). An explicit calculation of the $5pt,nonSSC$ term could be needed before robust conclusions can be drawn on this point. The inclusion of galaxy shape-noise and shot-noise (which are diagonal) in realistic lensing and galaxy clustering applications is also expected to stabilize the inversion of covariance matrices with incomplete off-diagonal contributions.

The derivation presented here can serve as the backbone for the calculation of the covariance matrix of squeezed galaxy bispectra, which is a relevant observable in studies of primordial non-Gaussianity of the local type \cite{scoccimarro/etal:2004, liguori/etal:2010, 2010CQGra..27l4011D, 2010AdAst2010E..64V, 2001PhRvD..63f3002K, jeong/komatsu:2009b, 2007PhRvD..76h3004S, 2010JCAP...07..002N, 2011JCAP...04..006B, sefusatti/etal:2012, 2016JCAP...06..014T, 2018arXiv180206762D, 2014JCAP...03..032T}. For that, the calculation should be extended to incorporate galaxy bias and redshift-space distortion effects \cite{2018arXiv180604015D}, as well as galaxy power spectrum responses \cite{2018JCAP...02..022L}. The calculation of the squeezed lensing bispectrum covariance can instead be readily obtained from that presented in this paper by performing the appropriate projections along the line-of-sight \cite{2005PhRvD..72h3001D, 2001ApJ...548....7C}. The weak-lensing bispectrum is however not very sensitive to primordial non-Gaussianity \cite{shearfNL}, but contains other cosmological information \cite{2013arXiv1306.4684K, 2018arXiv181002374C, 2018arXiv181207437R}. Another practical application is in cross-checks of $N$-body ensemble estimates of the general bispectrum covariance (e.g.~Refs.~\cite{2006PhRvD..74b3522S, 2013MNRAS.429..344K, 2017PhRvD..96b3528C, 2018PhRvD..97d3532C, 2017MNRAS.465.1757G, 2018arXiv180302132S, 2018arXiv180609499C}), which normally require a large number of simulations to have statistical errors under control. Our calculation can then serve as a useful point of comparison in the squeezed limit, where it is effectively complete and noise-free. 

Finally, we end by noting that the response approach can straightforwardly be extended to include bispectrum response functions, which can also be measured with separate universe simulations. These developments, which are left for future work, will permit generalizing the derivation presented here beyond squeezed bispectrum configurations.

\begin{acknowledgments}
We thank Fabian Schmidt for his invaluable input and insights throughout this work. We would also like to thank Linda Blot, Olivier Dor\'e, Elisabeth Krause, Fabien Lacasa, Roland de Putter, Matteo Rizzato and Shun Saito for many useful comments and conversations.
\end{acknowledgments}

\appendix 

\section{Diagram rules for cosmological perturbation theory}\label{app:feynman}

This appendix displays the diagram rules we adopt to compute $n$-point functions in cosmological perturbation theory. The conventions are based on those of Ref.~\cite{abolhasani/mirbabayi/pajer:2016}. The rules are as follows: 
\begin{enumerate}
\item An $n$-point correlation function is represented by a number of diagrams with $n$ outgoing external legs. 
\item Interaction vertices have $m\geq 2$ ingoing lines $\vp_1,\cdots,\vp_m$ that couple to a single outgoing line $\vp$; each vertex is assigned a factor
  \be
   m! F_m(\vp_1, \cdots, \vp_m) (2\pi)^3 \d_D(\vp-\vp_{1\cdots m})\,.
  \ee
Ingoing and outgoing momenta are assigned a negative and positive sign, respectively. Each ingoing line must be directly connected to a propagator (linear power spectrum). The $F_m$ are called symmetrized perturbation theory kernels \cite{Bernardeau/etal:2002}.

\item Propagators are represented as vertices with 2 outgoing lines of equal momentum $p$ as {\includegraphicsbox[scale=0.8]{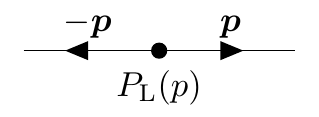}}; each is assigned a factor $\Plin(p)$. To ease the notation, we often skip labeling the propagator outgoing lines (which line is which can be inferred from momentum conservation).

\item All momenta that are not fixed in terms of momentum constraints are called loop momenta and are integrated over as
  \be
  \int \frac{d^3\vv}{(2\pi)^3}\,.
  \ee
A tree-level diagram is a diagram without any such loop integrals.
\item Each diagram is multiplied by the {\it symmetry factor}, which accounts for degenerate configurations of the diagram, as well as all nonequivalent labellings of external lines.
\end{enumerate}

\noindent The inclusion of response-type interactions can be achieved with one additional rule:

\begin{enumerate}\setcounter{enumi}{5}
\item Response-type vertices, $\R_n$, have 2 (instead of 1) outgoing lines with momenta $\vp,\vq$, and $n\geq 1$ incoming lines with momenta $\vr_a$. These vertices are only predictive in the limit where $\sum_a r_a \ll \min\{p, q, \knl\}$, but no restriction is placed on the magnitude of the outgoing momenta, which can be in the nonlinear regime. In our notation, we represent them as dashed blobs. Each such vertex is assigned a factor
  \be
 \frac12 \R_n(p; \cdots)\, P_m(p) (2\pi)^3 \d_D(\vp+\vq - \vr_{1\cdots a})\,;
  \ee
the dots in the argument of $\R_n$ denote all the relevant cosine angles and soft momenta magnitude ratios that exist at a given order $n$ (described in detail in Sec.~2 of Ref.~\cite{responses1}). The factor $1/2$ cancels the trivial permutation $\vp \leftrightarrow \vq$, which is always present when response vertices appear.
\end{enumerate}

\section{Permutations in $\cov^{BB}_{BB}$, $\cov^{BB}_{TP}$ and $\cov^{BP}_{BP}$}\label{app:BBTPperms}

All the permutations of the $BB$ and $TP$ contributions to the squeezed-bispectrum covariance can be written as
\bq\label{eq:perms_1}
\cov^{BB}_{BB, TP} &=& \frac{(2\pi)^3}{V_WV_{k_1k_1's_1}V_{k_2k_2's_2}} \int_{k_1}{\rm d}^3\vp\int_{k_1'}{\rm d}^3\vq \int_{s_1}{\rm d}^3\vr \int_{k_2}{\rm d}^3\vp'\int_{k_2'}{\rm d}^3\vq' \int_{s_2}{\rm d}^3\vr' \nonumber \\
&& \delta_D(\vp+\vq+\vr)\delta_D(\vp'+\vq'+\vr') \Big[A + B + C + D + E + F + G + H + I\Big], \nonumber \\
\eq
where the terms inside square brackets are given by
\bq\label{eq:perms_2}
A_{BB} &=&  \delta_{s_1s_2} B_m(\vp, \vq, \vr') B_m(\vp', \vq', \vr) \delta_D(\vr-\vr')       ; \ A_{TP} = \delta_{s_1s_2} P_m(r)T_m(\vp, \vq, \vp', \vq') \delta_D(\vr+\vr'), \nonumber \\
B_{BB} &= & \delta_{k_1'k_2}  B_m(\vp, \vp', \vr) B_m(\vq, \vq', \vr') \delta_D(\vq-\vp')   ; \ B_{TP} = \delta_{k_1'k_2} P_m(q)T_m(\vp, \vq', \vr, \vr') \delta_D(\vq+\vp'), \nonumber \\
C_{BB} &= & \delta_{k_1'k_2'}  B_m(\vp, \vq', \vr) B_m(\vp', \vq, \vr') \delta_D(\vq-\vq')  ;  \ C_{TP} = \delta_{k_1'k_2'} P_m(q)T_m(\vp, \vp', \vr, \vr') \delta_D(\vq+\vq'), \nonumber \\
D_{BB} &= & \delta_{k_1k_2'}  B_m(\vp, \vp', \vr') B_m(\vq, \vq', \vr) \delta_D(\vp-\vq')   ; \ D_{TP} = \delta_{k_1k_2'} P_m(p)T_m(\vp', \vq, \vr, \vr') \delta_D(\vp+\vq'), \nonumber \\
E_{BB} &= & \delta_{k_1k_2} B_m(\vp, \vq', \vr') B_m(\vp', \vq, \vr)  \delta_D(\vp-\vp')    ;  \ E_{TP} = \delta_{k_1k_2} P_m(p)T_m(\vq, \vq', \vr, \vr') \delta_D(\vp+\vp'), \nonumber \\
F_{BB} &= & \delta_{k_1s_2}  B_m(\vp, \vp', \vq') B_m(\vq, \vr, \vr')  \delta_D(\vp-\vr')    ;  \ F_{TP} =  \delta_{k_1s_2} P_m(p)T_m(\vp', \vq', \vq, \vr) \delta_D(\vp+\vr'), \nonumber \\
G_{BB} &=& \delta_{k_1's_2} B_m(\vq, \vp', \vq') B_m(\vp, \vr, \vr')  \delta_D(\vq-\vr')    ;  \ G_{TP} =  \delta_{k_1's_2} P_m(q)T_m(\vp', \vq', \vp, \vr) \delta_D(\vq+\vr'), \nonumber \\
H_{BB} &=& \delta_{k_2s_1} B_m(\vp, \vq, \vp') B_m(\vq', \vr, \vr')  \delta_D(\vp'-\vr)     ;  \ H_{TP} =  \delta_{k_2s_1} P_m(r)T_m(\vp, \vq, \vq', \vr') \delta_D(\vp'+\vr), \nonumber \\
I_{BB}  &=& \delta_{k_2's_1}  B_m(\vp, \vq, \vq') B_m(\vp', \vr, \vr') \delta_D(\vq'-\vr)     ;  \  I_{TP} =  \delta_{k_2's_1} P_m(r)T_m(\vp, \vq, \vp', \vr') \delta_D(\vq'+\vr), \nonumber \\
\eq
for the $BB$ and $TP$ contributions, as indicated by the subscripts. We skipped writing the permutation of the $BB$ term that cancels with the $B_m(k_1, k_1', s_1)B_m(k_2, k_2', s_2)$  term in Eq.~(\ref{eq:covBB_def}).

All the permutations of the $BP$ term of the squeezed-bispectrum and power spectrum cross-covariance can be written as

\bq\label{eq:perms_3}
\cov^{BP}_{BP} &=& \frac{(2\pi)^3}{V_WV_{k_1k_1's_1}V_{k_2}} \int_{k_1}{\rm d}^3\vp\int_{k_1'}{\rm d}^3\vq \int_{s_1}{\rm d}^3\vr \int_{k_2}{\rm d}^3\vp' \delta_D(\vp+\vq+\vr) \nonumber \\
&& \Big[B_0 + B_1 + E_0 + E_1 + H_0 + H_1\Big], \nonumber \\
\eq
where
\bq\label{eq:perms_4}
B_0 &=& \delta_{k_1'k_2} \Pnl(q) B(\vp, \vq, \vr) \delta_D(\vq + \vp') \ \ \ ; \ \ \  B_1 = \delta_{k_1'k_2} \Pnl(q) B(\vp, \vq, \vr) \delta_D(\vq - \vp') \nonumber \\
E_0 &=& \delta_{k_1k_2} \Pnl(p) B(\vp, \vq, \vr) \delta_D(\vp + \vp') \ \ \ ; \ \ \  E_1 = \delta_{k_1k_2} \Pnl(p) B(\vp, \vq, \vr) \delta_D(\vp - \vp') \nonumber \\
H_0 &=& \delta_{s_1k_2} \Pnl(r) B(\vp, \vq, \vr) \delta_D(\vr + \vp') \ \ \ ; \ \ \  H_1 = \delta_{s_1k_2} \Pnl(r) B(\vp, \vq, \vr) \delta_D(\vr - \vp').
\eq
We skipped writing explicitly the $BP$ permutation that cancels with the $B_m(k_1, k_1', s_1)\Pnl(k_2)$  term in Eq.~(\ref{eq:covBP_def}).

\section{Derivation of the SSC term of the squeezed matter bispectrum, $Q_{m,6}^{SSC}$}\label{app:SSC_BB}

In this appendix, we derive the expression of $Q_{m,6}^{SSC}$ that determines the SSC contribution of the squeezed bispectrum covariance (cf.~Eq.~(\ref{eq:6pt_5})). The polyspectra $Q_{m,6}^{SSC}$ associated with the connected 6-point function is defined as
\bq\label{eq:app_qm6}
(2\pi)^3 Q_{m,6}(\vk_a, \vk_b, \vk_c, \vk_d, \vk_e, \vk_f) \delta_D(\vk_{abcdef})= \big< \tilde{\delta}(\vk_a)\tilde{\delta}(\vk_b)\tilde{\delta}(\vk_c)\tilde{\delta}(\vk_d)\tilde{\delta}(\vk_e)\tilde{\delta}(\vk_f)\big>_c,
\eq
which at tree level in perturbation theory is determined by the permutations of the following six types of diagrams:
\bq\label{eq:Qm6_diag1}
Q_{m,6}^{F_3^2} &=& \left({\raisebox{0.0cm}{\includegraphicsbox[scale=0.9]{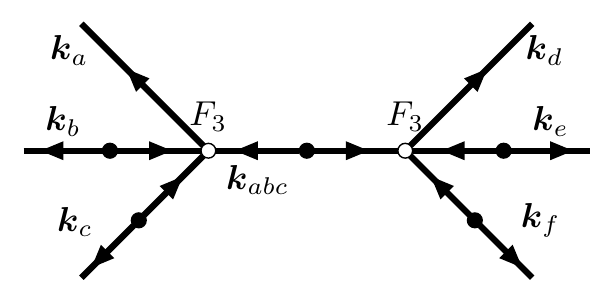}}} + {\rm permutations}\right) \nonumber \\
&=& (3!)^2F_3(\vk_b, \vk_c, -\vk_{abc}) F_3(\vk_e, \vk_f, \vk_{abc}) \Plin(k_b)\Plin(k_c)\Plin(k_e)\Plin(k_f)\Plin(|\vk_{abc}|) \nonumber \\
 &+& {\rm permutations},
\eq
\bq\label{eq:Qm6_diag2}
Q_{m,6}^{F_2^4} &=& \left({\raisebox{0.0cm}{\includegraphicsbox[scale=0.9]{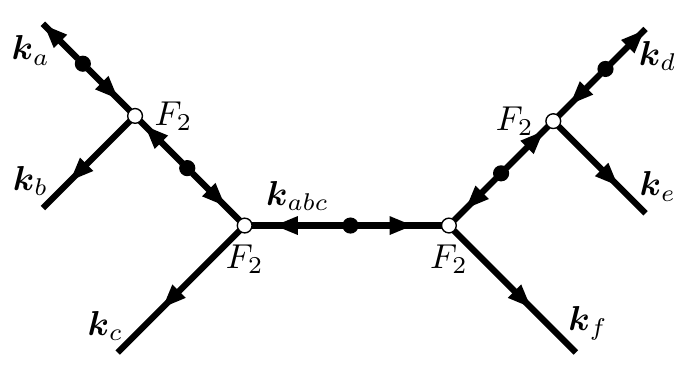}}} + {\rm permutations}\right) \nonumber \\
&=& (2!)^4F_2(\vk_a, -\vk_{ab}) F_2(\vk_{ab}, -\vk_{abc}) F_2(\vk_d, -\vk_{ed}) F_2(\vk_{ed}, \vk_{abc}) \Plin(k_a)\Plin(k_d) \Plin(|\vk_{ab}|) \nonumber \\
&\times&\Plin(|\vk_{ed}|) \Plin(|\vk_{abc}|) + {\rm permutations},
\eq
\bq\label{eq:Qm6_diag5}
Q_{m,6}^{F_2^2F_3(A)} &=& \left({\raisebox{0.0cm}{\includegraphicsbox[scale=0.9]{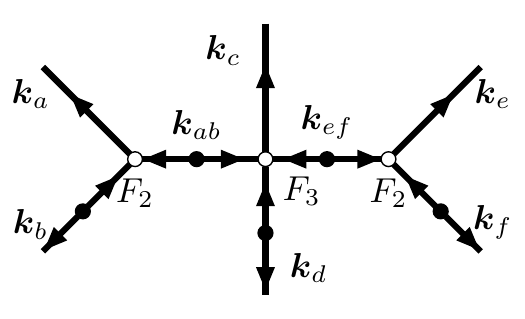}}} + {\rm permutations}\right) \nonumber \\
&=&  3!(2!)^2F_2(\vk_b, -\vk_{ab}) F_3(\vk_d, \vk_{ab}, \vk_{ef})F_2(\vk_f, -\vk_{ef}) \nonumber \\
&&\times \Plin(k_b)\Plin(k_d)\Plin(k_f)\Plin(|\vk_{ab}|)\Plin(|\vk_{ef}|) + {\rm permutations},
\eq
\bq\label{eq:Qm6_diag3}
Q_{m,6}^{F_2^2F_3(B)} &=& \left({\raisebox{0.0cm}{\includegraphicsbox[scale=0.9]{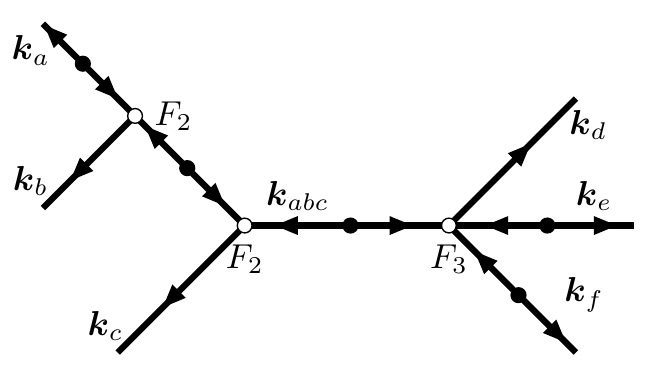}}} + {\rm permutations}\right) \nonumber \\
&=& 3!(2!)^2 F_2(\vk_a, -\vk_{ab}) F_2(\vk_{ab}, -\vk_{abc}) F_3(\vk_e, \vk_f, \vk_{abc}) \nonumber \\
&& \times \Plin(k_a)\Plin(k_e) \Plin(k_f) \Plin(|\vk_{ab}|) \Plin(|\vk_{abc}|) + {\rm permutations},
\eq
\bq\label{eq:Qm6_diag4}
Q_{m,6}^{F_5} &=& \left({\raisebox{0.0cm}{\includegraphicsbox[scale=0.9]{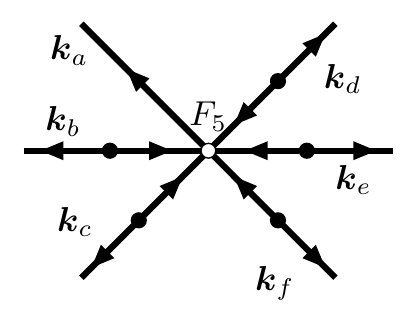}}} + {\rm permutations}\right) \nonumber \\
&=& 5! F_5(\vk_b, \vk_c, \vk_d, \vk_e, \vk_f) \Plin(k_b) \Plin(k_c) \Plin(k_d) \Plin(k_e) \Plin(k_f)  + {\rm permutations},
\eq
\bq\label{eq:Qm6_diag6}
Q_{m,6}^{F_2F_4} &=& \left({\raisebox{0.0cm}{\includegraphicsbox[scale=0.9]{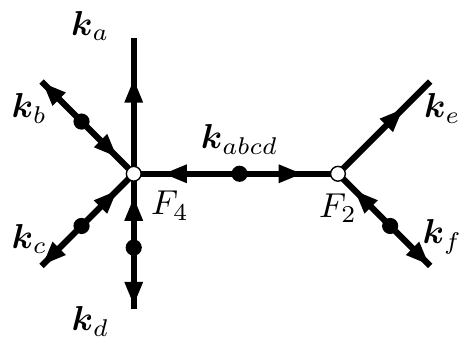}}} + {\rm permutations}\right) \nonumber \\
&=& 4!2!F_4(\vk_b, \vk_c, \vk_d, -\vk_{abcd}) F_2(\vk_f, \vk_{abcd}) \Plin(k_b) \Plin(k_c) \Plin(k_d) \Plin(k_f) \Plin(|\vk_{abcd}|) \nonumber \\ 
 &+& {\rm permutations}.
\eq
The connected 6-point function contribution is determined by the configuration $Q_{m,6}^{SSC} = Q_{m,6}(\vp, \vq, \vr + \vv, \vp', \vq', \vr' - \vv)$ and the corresponding SSC piece corresponds to all of the permutations that vanish for $v=0$. Inspecting the above diagrams, one notes that the SSC terms must be diagrams that contain lines that propagate momenta $\vk_{abc}$, which are the diagrams $Q_{m,6}^{F_3^2}$, $Q_{m,6}^{F_2^4}$ and $Q_{m,6}^{F_2^2F_3(B)}$. More specifically, the SSC term is determined by the permutations of these diagrams that yield $\vk_{abc} = \vp + \vq + \vr + \vv = \vv$, which are thus proportional to $\Plin(v)$ and vanish for $v=0$ (we have implicitly used the momenta constrain $\delta_D(\vp + \vq + \vr)$ that enters in Eq.~(\ref{eq:6pt_5})).

Explicitly, at tree level in perturbation theory, the SSC part of the bispectrum covariance is determined by
\bq\label{eq:Qm6_ssc_tree}
Q_{m,6}^{SSC}(\vp, \vq, \vr, \vp', \vq', \vr' | \vv) & = & {\raisebox{0.0cm}{\includegraphicsbox[scale=0.9]{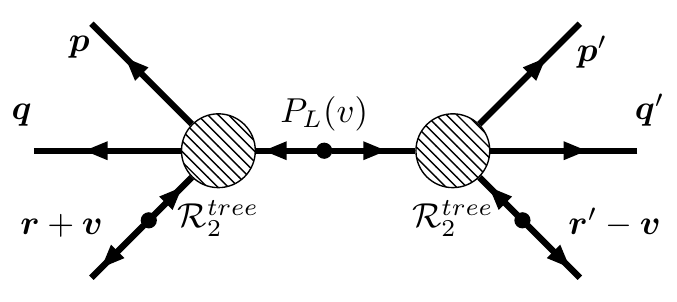}}} \nonumber \\
&=& \R_2^{\rm tree}(p, \mu_{\vp,\vr}, \mu_{\vp,\vv}, \mu_{\vr,\vv}, r/v) \R_2^{\rm tree}(p', \mu_{\vp',\vr'}, -\mu_{\vp',\vv}, -\mu_{\vr',\vv}, r'/v) \nonumber \\
&\times& \Plin(p)\Plin(p')\Plin(r)\Plin(r')\Plin(v) \nonumber \\
&=& \left[\left({\raisebox{0.0cm}{\includegraphicsbox[scale=0.9]{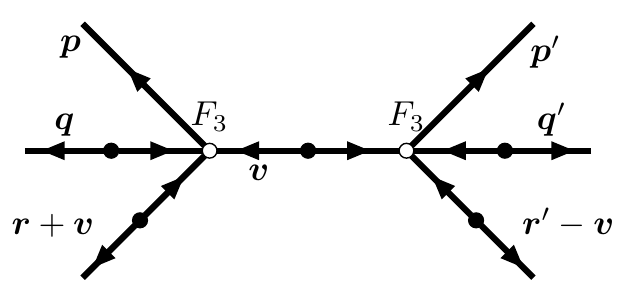}}} + (\vp \leftrightarrow \vq)\right)  (\vp' \leftrightarrow \vq') \right] \nonumber \\
&+& \left[\left({\raisebox{0.0cm}{\includegraphicsbox[scale=0.9]{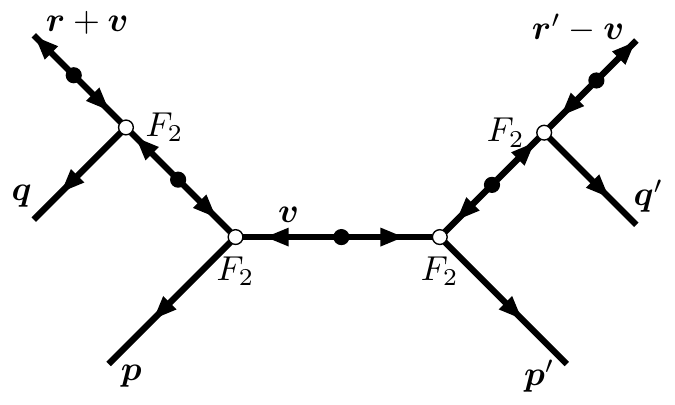}}} + (\vp \leftrightarrow \vq)\right)  (\vp' \leftrightarrow \vq') \right] \nonumber \\
&+& \left[\left({\raisebox{0.0cm}{\includegraphicsbox[scale=0.9]{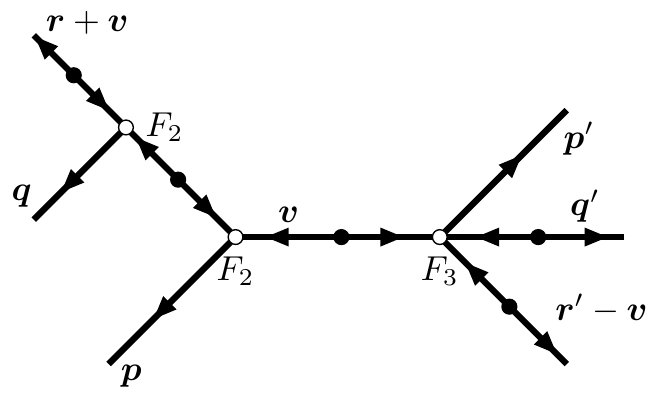}}} + (\vp \leftrightarrow \vq)\right)  (\vp' \leftrightarrow \vq') \right] \nonumber \\ 
&+& \left[\left({\raisebox{0.0cm}{\includegraphicsbox[scale=0.9]{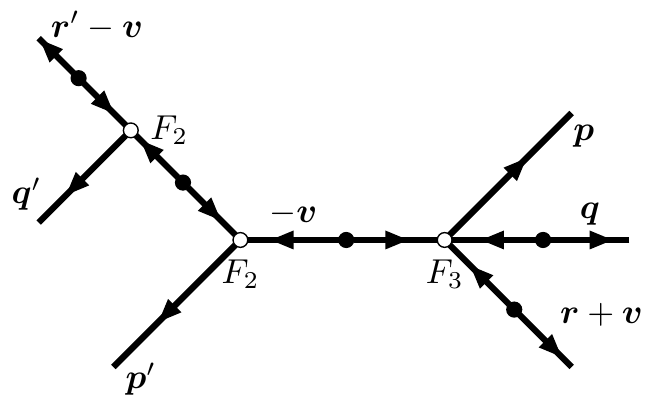}}} + (\vp \leftrightarrow \vq)\right)  (\vp' \leftrightarrow \vq') \right]. \nonumber \\ 
\eq
In terms of perturbation theory kernels, the above can be written as
\bq\label{eq:Qm6_ssc_tree2}
Q_{m,6}^{SSC} &=& \Bigg[ (3!)^2F_3(\vq, \vr+\vv, -\vv) F_3(\vq', \vr'-\vv, \vv) \Plin(q)\Plin(|\vr + \vv|)\Plin(q')\Plin(|\vr' - \vv|)\Plin(v) \nonumber \\
&+& (2!)^4F_2(-\vq-\vr-\vv,\vr + \vv) F_2(\vq+\vr+\vv, -\vv) \nonumber \\
&&\ \ \ \ \ \ \ \ \ \ \ \ \ \ \ \ \ \ \ \  \times F_2(-\vq'-\vr'+\vv,\vr' - \vv) F_2(\vq'+\vr'-\vv, \vv) \nonumber \\
&&\ \ \ \ \ \ \ \ \ \ \ \ \ \ \ \ \ \ \ \  \times \Plin(|\vr+\vv|)\Plin(|\vq+\vr+\vv|) \Plin(|\vr' - \vv|) \Plin(|\vq' + \vr' - \vv|) \Plin(v) \nonumber \\
&& \nonumber \\
&+& 3!(2!)^2F_2(-\vq-\vr-\vv,\vr + \vv) F_2(\vq+\vr+\vv, -\vv) F_3(\vq', \vr-\vv, \vv) \nonumber \\
&&\ \ \ \ \ \ \ \ \ \ \ \ \ \ \ \ \ \ \ \  \times \Plin(|\vr+\vv|)\Plin(|\vq+\vr+\vv|) \Plin(q') \Plin(|\vr' - \vv|) \Plin(v) \nonumber \\
&& \nonumber \\
&+& 3!(2!)^2F_2(-\vq'-\vr'+\vv,\vr' - \vv) F_2(\vq'+\vr'-\vv, \vv) F_3(\vq, \vr+\vv, -\vv) \nonumber \\ 
&&\ \ \ \ \ \ \ \ \ \ \ \ \ \ \ \ \ \ \ \  \times \Plin(|\vr'-\vv|)\Plin(|\vq'+\vr' - \vv|) \Plin(|\vr+\vv|) \Plin(q) \Plin(v) \nonumber \\
&+& (\vp \leftrightarrow \vq)\Bigg]  + (\vp' \leftrightarrow \vq').
\eq
The above expressions are only strictly valid in the linear regime of all modes involved. However, by replacing the tree-level $\R_2^{\rm tree}$ vertices with the resummed vertices measured from separate universe simulations $\R_2$ (cf.~Sec.~\ref{sec:responses}), as well as replacing the linear power spectrum of the hard triangle modes $\vp, \vq, \vp', \vq'$ by the fully evolved nonlinear power spectrum, then the result becomes valid for linear values of the modes $\vr, \vr', \vv$, but any nonlinear value of $\vp, \vq, \vp', \vq'$. This finalizes our derivation of Eq.~(\ref{eq:6pt_3}).

The derivation of the SSC contribution to the cross-covariance $\cov^{BP}$ is in all analogous to that described in this appendix, and hence we skip showing it explicitly  (see also Appendix A of Ref.~\cite{akitsu/takada/li} or Appendix B of Ref.~\cite{completessc} for the derivation of the tree-level power spectrum SSC term).

\section{On the numerical evaluation of the integrals}\label{app:numericalevaluation}
The numerical evaluation of the covariance contributions listed in the main body of the paper involves performing a large number of bin-average integrals with constraints amongst the integrated momenta. In this appendix, we describe a simple Monte Carlo numerical recipe that we use to evaluate those integrals.

Let us assume we are interested in the following integral
\bq\label{eq:I1_1}
I_1(k_1, k_1', s_1) &=&  \int_{k_1}{\rm d}^3\vp\int_{k_1'}{\rm d}^3\vq \int_{s_1}{\rm d}^3\vr\ \delta_D(\vp+\vq+\vr) f(\vp, \vq, \vr),
\eq
which can be evaluated with Monte Carlo integration as
\bq\label{eq:I1_2}
I_1(k_1, k_1', s_1) &\approx& \big<f(\vp, \vq, \vr)\big>_{MC} \int_{k_1}{\rm d}^3\vp\int_{k_1'}{\rm d}^3\vq \int_{s_1}{\rm d}^3\vr\ \delta_D(\vp+\vq+\vr) \nonumber \\
&=& \big<f(\vp, \vq, \vr)\big>_{MC} V_{k_1k_1's_1},
\eq
where $V_{k_1k_1's_1}$ is the integration volume and $\big<f(\vp, \vq, \vr)\big>_{MC}$ is the average of the integrand $f$ over a number of momenta samples drawn from each wavenumber shell and that satisfy the constraint $\delta_D(\vp+\vq+\vr)$. These samples can be obtained as follows:
\begin{enumerate}

\item Uniformly sample $\vr$ in the $s_1$ shell, i.e., $r \draw U(s_1-\Delta s_1/2, s_1+\Delta s_1/2)$, $\theta_{\vr} \draw U(0, \pi)$ and $\varphi_{\vr} \draw U(0, 2\pi)$;

\item Uniformly sample $\vp$ in the $k_1$ shell, i.e., $p \draw U(k_1-\Delta k_1/2, k_1+\Delta k_1/2)$, $\theta_{\vp} \draw U(0, \pi)$ and $\varphi_{\vp} \draw U(0, 2\pi)$;

\item Compute $\vq = -\vp - \vr$. If $q \in \left[k_1' - \Delta k_1'/2; k_1' + \Delta k_1'/2\right]$ then one accepts the three vectors, otherwise one goes back to point 1.

\end{enumerate}

The $BB$ and $TP$ contributions to $\cov^{BB}$ are, however, given by integrals of the form
\bq\label{eq:I2_1}
I_2(k_1, k_1', s_1, k_2, k_2', s_2) &=&  \int_{k_1}{\rm d}^3\vp\int_{k_1'}{\rm d}^3\vq \int_{s_1}{\rm d}^3\vr \int_{k_2}{\rm d}^3\vp'\int_{k_2'}{\rm d}^3\vq' \int_{s_2}{\rm d}^3\vr'\  \nonumber \\
&& \delta_D(\vp+\vq+\vr) \delta_D(\vp'+\vq'+\vr') \delta_D(\vr - \vr') f(\vp, \vq, \vr, \vp', \vq', \vr'), \nonumber \\
\eq
i.e., two momentum constraints on $\{\vp, \vq, \vr\}$ and $\{\vp', \vq', \vr'\}$, and another constraint linking momenta of these two sets, which in this case is $\vr = \vr'$. With Monte Carlo integration we can evaluate this term as
\bq\label{eq:I2_2}
I_2(k_1, k_1', s_1, k_2, k_2', s_2) &\approx&  \big<f(\vp, \vq, \vr, \vp', \vq', \vr')\big>_{MC} \nonumber \\ 
&& \int_{k_1}{\rm d}^3\vp\int_{k_1'}{\rm d}^3\vq \int_{s_1}{\rm d}^3\vr \int_{k_2}{\rm d}^3\vp'\int_{k_2'}{\rm d}^3\vq' \int_{s_2}{\rm d}^3\vr'\ \nonumber \\
&& \delta_D(\vp+\vq+\vr) \delta_D(\vp'+\vq'+\vr') \delta_D(\vr - \vr'), \nonumber \\ 
&=& \big<f(\vp, \vq, \vr, \vp', \vq', \vr')\big>_{MC} U(s_1, s_2),
\eq
where $U(s_1, s_2) = \delta_{s_1s_2} 16\pi^3k_1k_2k_1'k_2' \Delta k_1\Delta k_1'\Delta k_2\Delta k_2'\Delta s_1$ is the integration volume (see Appendix B of Ref.~\cite{2017PhRvD..96b3528C} for a derivation). The sampling of the function $f$ can be done by first sampling the set $\{\vp, \vq, \vr\}$ with the recipe above. The set $\{\vp', \vq', \vr'\}$ is then subsequently sampled using essentially the same recipe, but instead of sampling $\vr'$ in point 1, one simply takes it to be $\vr' = \vr$. The integration of all other permutations of the $\cov^{BB}_{BB}$ and $\cov^{BB}_{TP}$ terms is carried out in a similar way.

The calculation of $\cov^{BB}_{6pt, SSC}$ involves integrals with momentum constraints $\delta_D(\vp + \vq + \vr)\delta_D(\vp' + \vq' + \vr')$ and can be performed by simply executing the recipe above twice (in addition to sampling also the window momenta $\vv$; cf.~Eq.~(\ref{eq:6pt_5})).

For the bin widths that we adopt in our numerical results, we found this straightforward sampling recipe to yield an acceptance rate between $50\%$ to $90\%$ (depending on the size of the bin-widths, which is not constant in a linear scale). For narrower bin-widths, the acceptance rate is expected to decrease. When the acceptance rate becomes too low, a possible way out would be to use the constrained sampling recipe outlined in Ref.~\cite{2017PhRvD..96b3528C}, which makes use of the fact that, for sufficiently narrow bins, the internal triangle $\{\vp, \vq, \vr\}$ is obtained by slight perturbations of the angles of the external triangle $\{k_1, k_1', s_1\}$.

\bibliography{REFS}

\end{document}